\documentclass[12pt,a4paper,final]{iopart}
\usepackage{iopams}

\expandafter\let\csname equation*\endcsname\relax
\expandafter\let\csname endequation*\endcsname\relax

\usepackage[utf8]{inputenc}
\usepackage{amsmath}
\usepackage{amsfonts}
\usepackage{amssymb}
\usepackage{amsthm}
\usepackage{graphicx}
\usepackage{bm}
\usepackage{bbm}
\usepackage{cite}
\usepackage{subcaption}
\usepackage{xcolor}   
\usepackage{soul}     

\usepackage{algorithm}
\usepackage{algpseudocode}

\usepackage[breaklinks=true,colorlinks=true,linkcolor=blue,urlcolor=blue,citecolor=blue]{hyperref}

\newcommand{\defeq}{\mathrel{\mathop:}=}

\renewcommand{\S}{\mathcal{S}}

\newcommand{\vth}{v_{\text{th}}}

\newcommand{\alphag}{\tilde{\alpha}}
\renewcommand{\thetag}{\tilde{\theta}}

\begin{document}
	
	\title[EM algorithm for kappa distributions]%
	{Parameter estimation for kappa distributions using the EM algorithm in the superstatistical framework}
	
	\author{Leonardo Herrera-Fuenzalida$^{1}$\footnote[1]{Corresponding author.} and Sergio Davis$^{2,1}$}
	
	\address{$^1$Departamento de Física y Astronomía, Facultad de Ciencias Exactas, Universidad Andrés Bello, Sazié 2212, piso 7, 8370136, Santiago, Chile}
	\address{$^2$Research Center in the Intersection in Plasma Physics, Matter and Complexity (P$^2$mc), Comisión Chilena de Energía Nuclear, Casilla 188-D, Santiago, Chile}
	
	\ead{l.herrerafuenzalida@uandresbello.edu}
	\ead{sergio.davis@cchen.cl}

\begin{abstract}
Kappa distributions are widely used in space plasma physics to model velocity distribution functions with heavy tails. Parameter estimation in these distributions is, however, complicated by the fact that the kappa distribution does not belong to the exponential family, so it admits no sufficient statistics and direct maximum likelihood requires numerical optimization without analytically closed-form update equations. Working within the Beck-Cohen superstatistics framework, where a gamma-distributed inverse temperature \(\beta\) generates the kappa distribution upon marginalization, we treat \(\beta\) as a latent variable. This hierarchical description restores the exponential family structure that the marginal kappa distribution lacks, and yields an analytically tractable implementation of the expectation-maximization (EM) algorithm whose E-step and M-step admit closed-form expressions in terms of sufficient statistics. Applied to synthetic data drawn from the model, the algorithm converges monotonically to a stationary point of the marginal kappa log-likelihood and recovers the generating parameters consistently across the explored range of \(\kappa\). EM thus offers a tractable and transparent route to inference in superstatistical systems with local temperature fluctuations.

\end{abstract}


\section{Introduction}

Research in collisionless plasmas frequently deals with velocity distributions that deviate from Maxwellian equilibrium and are successfully modeled by the \textit{kappa} distribution \cite{Binsack1966, Olbert1968, Vasyliunas1968}. This distribution has been reported not only in space plasmas but also in laboratory plasmas, as documented in recent work \cite{Nicolaou2022}. However, a methodologically rigorous approach to parameter inference in these distributions remains underdeveloped. A fundamental question persists: how can we robustly estimate the \(\kappa\) (kappa) parameter once we assume that the data follow this distribution?

In order to address this question, it is necessary to ground the problem within appropriate statistical physics frameworks that account for the origin of heavy-tailed distributions. 
Two frameworks in particular stand out in the literature, namely Tsallis' non-extensive statistics~\cite{Tsallis1988} and Beck and Cohen's superstatistics~\cite{Beck2003, Beck2004}. The probabilistic structure provided by superstatistics offers practical pathways for estimating the \(\kappa\) parameter, which we develop in this work through a hierarchical statistical approach based on the Expectation-Maximization algorithm (EM for short)~\cite{Dempster1977}.

Superstatistics provides a probabilistic framework for modeling non-equilibrium systems based on the idea that intensive parameters, such as the inverse temperature $\beta \defeq 1/(k_B T)$, fluctuate. Within this formalism, the microstate probability distribution is obtained by marginalizing over the joint distribution of microstates and inverse temperatures \(\beta\). Originally formalized by Beck and Cohen~\cite{Beck2003, Beck2004}, superstatistics has proven effective across plasma physics~\cite{Sanchez2021, Ourabah2015, Davis2019, Ourabah2020, Gravanis2021}, condensed matter systems~\cite{Dixit2013, Dixit2015, Herron2021}, high-energy physics and cosmology~\cite{Jizba2010, Ayala2018, Ourabah2019}, and other domains~\cite{Chen2008, Denys2016, Bogachev2017, Schaefer2018, Costa2022, Sanchez2025}. The key advantage of the superstatistical approach is its agnosticism, in the sense that the framework does not require specification of the underlying physical mechanism generating the fluctuations of \(\beta\), only that such fluctuations exist and can be characterized probabilistically. This property makes superstatistics particularly suited to inference problems where the microscopic origin of this kind of heterogeneity is unknown or complex.

On the other hand, the EM algorithm, as presented by Dempster, Laird, and Rubin~\cite{Dempster1977}, operates in the space of probability distributions as a general framework, with particular utility for this work in hierarchical settings where the observed data \(\bm{X}\) follow a specified density \(P(\bm{X} \mid \bm{\lambda})\) and the parameters \(\bm{\lambda}\) are themselves characterized by a density \(P(\bm{\lambda} \mid \bm{\phi})\) governed by \emph{hyperparameters} \(\bm{\phi}\) at a deeper level of the model. Within this hierarchy, \(\bm{\phi}\) is estimated by maximizing the marginal likelihood, given by
\begin{equation}
	P(\bm{X}|\bm\phi) = \int_{\Lambda} d\bm\lambda \, P(\bm{X}|\bm\lambda)\,P(\bm\lambda|\bm\phi),
\end{equation}
where \(\bm\lambda\) plays the role of the latent variable (with \(\Lambda\) its support) and \(\bm\phi\) are the hyperparameters to be estimated.

This hierarchical structure is precisely the superstatistics setting: a parameter distributed with a density governed by hyperparameters. Here the inverse temperature \(\beta\) is the latent variable, and $\bm{\phi} = (\alpha, \theta)$, which govern \(P(\beta \mid \alpha, \theta)\), are the hyperparameters to estimate. 

As we will show in the following sections, the kappa distribution emerges naturally as the marginal density \(P(\bm{v}|\alpha, \theta)\) when \(\beta\) is gamma-distributed, that is, when
\begin{equation*}
 P(\beta|\alpha, \theta) = \frac{1}{\Gamma(\alpha)\theta^{\alpha}}\exp\left(-\frac{\beta}{\theta}\right)\beta^{\alpha-1}.
\end{equation*}
The hierarchical Gaussian-with-gamma-precision structure underlying our construction has a clear precedent antecedent in Section 4.6 of the original EM article \cite{Dempster1977}, where it appears without a physical interpretation; we make the connection precise at the end of Section 3 and develop our extension in Section 4.
%
\section{Superstatistics}
\label{sec:superstatistics}

%
%
%
\subsection{The concept of superstatistics}
The superstatistics framework~\cite{Beck2003,Beck2004} assumes that the inverse temperature \(\beta\) of a system is not fixed but has a probability of realization given by \(P(\beta \mid \bm{\phi})\). Given a particular value of \(\beta\), the states \(\bm{X}\) of the system are distributed according to a canonical distribution
\begin{equation}
	P(\bm{X} \mid \beta) \propto e^{-\beta H(\bm{X})},
\end{equation}
where \(H(\bm{X})\) is the Hamiltonian of the system. The marginal 
distribution of \(\bm{X}\) is thus obtained by integrating over all 
possible values of \(\beta\),
\begin{equation}
	P(\bm{X} \mid \bm{\phi}) = \int_{0}^{\infty} d\beta\, P(\bm{X} \mid \beta)\, 
	P(\beta \mid \bm{\phi}).
\end{equation}


%
%
%
\subsection{The kappa distribution from superstatistics}

In plasma physics, particle velocities $\bm V \defeq \{\bm v_1, \bm v_2, \bm v_3, \cdots, \bm v_N\}$ (or equivalently, their kinetic energies) are directly measurable quantities, whereas the inverse temperature $\beta$ is not directly observable. In this scenario, we can postulate a gamma distribution for the inverse temperature,
\begin{equation}
	\label{eq:gamma_distribution_of_beta}
	P(\beta|\alpha, \theta) = \frac{1}{\theta^\alpha \Gamma(\alpha)}\beta^{\alpha-1}\exp\left(-\frac{\beta}{\theta}\right),
\end{equation}
where $\alpha > 0$ and $\theta > 0$ are the shape and scale parameters, respectively. Without assuming equilibrium, the Maxwell-Boltzmann distribution given $\beta$ for a velocity datum $\bm v \in \mathbb{R}^d$, with $d$ the dimensionality of the velocity vector accessible to the diagnostic, is
\begin{equation}
	\label{eq:maxwell_distribution_of_velocities}
	P(\bm v |\beta) = \left(\frac{m\beta}{2\pi}\right)^{d/2} \exp\left(-\frac{\beta m \bm v^2}{2}\right).
\end{equation}
The marginal distribution $P(\bm v|\alpha, \theta)$, obtained by integrating over all possible values of $\beta$, is then given by
\begin{equation}
	\label{eq:marginal_super_kappa_gamma}
	P(\bm v|\alpha, \theta) = \int_0^{\infty} P(\beta| \alpha, \theta)\, P(\bm v|\beta)\, d\beta.
\end{equation}
Performing the integral, we obtain the marginal velocity distribution still expressed in terms of the parameters $\alpha$ and $\theta$ as
\begin{equation}
	\label{eq:kappa_marginal_alpha_theta}
	P(\bm v|\alpha, \theta) = \left(\frac{m \theta}{2\pi}\right)^{d/2}
	\frac{\Gamma(\alpha + d/2)}{\Gamma(\alpha)}
	\left(1+\frac{m \theta}{2}\bm v^2 \right)^{-(\alpha + d/2)}.
\end{equation}

\noindent
We now introduce the reparameterization that connects \eqref{eq:kappa_marginal_alpha_theta} with the kappa distribution as conventionally written in plasma physics~\cite{livadiotis2013}, namely
\begin{equation}
	\label{eq:reparam_kappa}
	\alpha + \frac{d}{2} \defeq \kappa + 1 \quad \text{and} \quad
	\frac{m \theta}{2} \defeq \frac{1}{(\kappa - d/2)\,v_{th}^2},
\end{equation}
where \(\kappa\) is the spectral index and $\vth$ the thermal velocity. This convention fixes the exponent of the marginal at \(-(\kappa - 1)\), no matter the dimension \(d\). So the same \(\kappa\) is obtained whether the data are one-dimensional (a single velocity component) or three-dimensional (the full velocity vector or its modulus). This choice brings expression \eqref{eq:kappa_marginal_alpha_theta} into the standard form
\begin{equation}
	\label{eq:kappa_standard}
	P(\bm v|\kappa, \vth) = \frac{1}{\eta_{\kappa}} \left[1+\frac{1}{\kappa - d/2}\frac{\bm v^2}{\vth^2}\right]^{-(\kappa+1)},
\end{equation}
where the normalization constant is
\begin{equation}
	\label{eq:normalization_constant_kappa_distribution}
	\eta_{\kappa} \defeq \frac{\Gamma(\kappa - d/2 + 1)}{\Gamma(\kappa + 1)}\left[\pi v_{th}^2 \left(\kappa - \tfrac{d}{2}\right)\right]^{d/2}.
\end{equation}
Setting $d = 3$ recovers the familiar three-dimensional kappa distribution with the well-known constraint $\kappa > 3/2$; the one-dimensional case ($d = 1$), relevant to single-component velocity diagnostics and to the implementation reported in Section~\ref{sec:results}, is recovered by setting $d = 1$, with the corresponding constraint $\kappa > 1/2$.

The kappa distribution exhibits the following key properties. For small $\kappa$ (but always $\kappa > d/2$ to ensure that the distribution is well-defined), the kappa distribution exhibits heavier tails than the Maxwellian, enabling it to capture \textit{rare events} and energetic particle populations. In the limit $\kappa \to \infty$, the kappa distribution converges to the Maxwell-Boltzmann distribution.

\subsection{Relation between the index $\kappa$ and the gamma distribution}

From the marginal in \eqref{eq:kappa_marginal_alpha_theta}, which retains the parameters $\alpha$ and $\theta$ of the gamma distribution of inverse temperature, the mean kinetic energy of a $d$-dimensional velocity vector is
\begin{equation}
	\label{eq:mean_kinetic_energy}
	\big<K\big>_{\alpha, \theta} = \left<\frac{m\bm v^2}{2}\right>_{\alpha, \theta} = \frac{d}{2\theta(\alpha-1)}, \qquad \alpha > 1.
\end{equation}
Combining \eqref{eq:reparam_kappa} and \eqref{eq:mean_kinetic_energy}, we can solve for $\alpha$ and $\theta$ in terms of $\kappa$, obtaining
\begin{subequations}
	\label{eq:relation_kappa_alpha_theta}
	\begin{align}
		\alpha & = \kappa + 1 - \frac{d}{2}, \\
		\theta & = \frac{d}{2\big<K\big>_{\alpha, \theta}\,(\kappa - d/2)},
	\end{align}
\end{subequations}
respectively. These relations allow us to determine $\kappa$ from the parameter $\alpha$ of the gamma distribution, with the inverse map
\begin{equation}
	\label{eq:kappa_from_alpha}
	\kappa = \alpha + \frac{d-2}{2}.
\end{equation}
For $d = 3$ this reduces to $\alpha = \kappa - 1/2$ and $\kappa = \alpha + 1/2$, while for $d = 1$ one has $\alpha = \kappa + 1/2$ and $\kappa = \alpha - 1/2$. As we shall see in Section~\ref{sec:EM_algorithm_SS}, the dimensionality $d$ enters the EM algorithm only through these two relations and through the posterior shape parameter $\tilde\alpha$ in the E-step, while the M-step update equations are themselves independent of $d$.


%
%
%
\subsection{Recent interpretations of superstatistics}

Recent developments in superstatistics have established that \(\beta\) is not a microscopic observable \cite{Davis2018}, supporting a Bayesian interpretation in which \(P(\beta|\bm{\phi})\) encodes a state 
of knowledge rather than a sampling distribution. Within this interpretation, an alternative parameterization of the gamma distribution in terms of the mean inverse temperature
\begin{equation}
\beta_S := \big<\beta\big>
\end{equation}
and the relative variance 
\begin{equation}
u := \frac{\big<(\delta\beta)^2\big>}{(\beta_S)^2}
\end{equation}
has been introduced~\cite{Davis2026}, related to the spectral index by \(\kappa = 1/u + 1/2\).

%
%
\section{Expectation-Maximization Algorithm}
\subsection{Maximum Likelihood Estimation as a Special Case of Maximum a Posteriori}
\label{sec:mle-as-map}

Given a set of \(N\) independent observations \(\bm{X} = \{\bm{x}_1, \bm{x}_2, \ldots, \bm{x}_N\}\) and a model \(P(\bm{x}_i \mid \bm{\lambda}, \mathcal{I})\) that assigns the probability of each datum conditional on the parameters \(\bm{\lambda}\) and the background information \(\mathcal{I}\), Bayes' theorem yields the posterior distribution of the parameters given the data,
\begin{equation}
	P(\bm{\lambda} \mid \bm{X}, \mathcal{I}) \;=\; \frac{P(\bm{X} \mid \bm{\lambda}, \mathcal{I}) \, P(\bm{\lambda} \mid \mathcal{I})}{P(\bm{X} \mid \mathcal{I})}.
	\label{eq:bayes-parameters}
\end{equation}
A natural point estimator within this Bayesian framework is the maximum a posteriori (MAP) estimator, defined as the value of \(\bm{\lambda}\) at which the posterior attains its maximum,
\begin{equation}
	\hat{\bm{\lambda}}_{\mathrm{MAP}} \;=\; \arg\max_{\bm{\lambda}} P(\bm{\lambda} \mid \bm{X}, \mathcal{I}).
	\label{eq:map-definition}
\end{equation}

In the absence of prior information that distinguishes any region of the parameter space, the prior \(P(\bm{\lambda} \mid \mathcal{I})\) is taken to be uniform over its domain. The posterior is then proportional to the sampling distribution of the data regarded as a function of the parameters,
\begin{equation}
	P(\bm{\lambda} \mid \bm{X}, \mathcal{I}) \;\propto\; P(\bm{X} \mid \bm{\lambda}, \mathcal{I}) \;\equiv\; \mathcal{L}(\bm{\lambda} \mid \bm{X}),
	\label{eq:posterior-as-likelihood}
\end{equation}
where \(\mathcal{L}(\bm{\lambda} \mid \bm{X})\) is known as the likelihood function. Under this flat-prior assumption the MAP estimator reduces to the classical maximum likelihood estimator (MLE),
\begin{equation}
	\hat{\bm{\lambda}}_{\mathrm{MLE}} \;=\; \arg\max_{\bm{\lambda}} \mathcal{L}(\bm{\lambda} \mid \bm{X}),
	\label{eq:mle-definition}
\end{equation}
which thus emerges as a particular case of Bayesian inference rather than as an independent principle.

Since the observations are assumed to be independent and identically distributed, the likelihood factorizes as
\begin{equation}
	\mathcal{L}(\bm{\lambda} \mid \bm{X}) \;=\; \prod_{i=1}^{N} P(\bm{x}_i \mid \bm{\lambda}, \mathcal{I}),
	\label{eq:likelihood-factorisation}
\end{equation}
and, because the logarithm is a monotonically increasing function, maximizing \(\mathcal{L}\) is equivalent to maximizing the log-likelihood,
\begin{equation}
	\ell(\bm{\lambda} \mid \bm{X}) \;=\; \sum_{i=1}^{N} \ln P(\bm{x}_i \mid \bm{\lambda}, \mathcal{I}),
	\label{eq:log-likelihood}
\end{equation}
which is computationally more convenient. The MLE is then obtained from
\begin{equation}
	\frac{\partial}{\partial \bm{\lambda}} \, \ell(\bm{\lambda} \mid \bm{X}) \;=\; 0.
	\label{eq:stationarity}
\end{equation}

\noindent
Substituting \eqref{eq:kappa_standard} into \eqref{eq:likelihood-factorisation}, we obtain the likelihood function for the kappa distribution as
\begin{equation}
\label{eq:verosimilitud_kappa}
\mathcal{L}(\kappa, v_{\text{th}}|\bm V) = \prod_{i=1}^{N} \frac{1}{\eta_{\kappa}} 
\left[1 + \frac{1}{\kappa - d/2} \frac{\bm v_i^2}{v^2_{\mathrm{th}}} 
\right]^{-(\kappa+1)},
\end{equation}
where the normalization factor $\eta_\kappa$ itself depends on $\kappa$ and $\vth$. Applying the extremum condition 
\[\frac{\partial}{\partial \kappa}\ln \mathcal{L} = 0\]
yields a transcendental equation with no closed-form solution, which must be solved numerically and is prone to instability or convergence to local maxima. Moreover, since the kappa distribution does not belong to the exponential family~\cite{bishop2006pattern}, it admits no sufficient statistics that would otherwise allow a compact, closed-form of the likelihood function~\cite{lehmann1998theory}.

These difficulties motivate an alternative approach grounded in the superstatistical paradigm: we recast the kappa distribution as a hierarchical model in which the observed velocities arise from a Maxwell-Boltzmann 
distribution conditioned on a gamma-distributed inverse temperature \(\beta\). Within this representation, \(\beta\) plays the role of a \emph{latent} (unobserved) variable, in such a way that the joint density of observed 
and latent quantities, known as the \emph{complete-data likelihood} \cite{Dempster1977}, does exhibit exponential family structure. The EM algorithm exploits precisely this property to perform maximum likelihood estimation 
iteratively, as developed in Section~\ref{sec:EM-Algorithm}.

%
%
%
%
\subsection{Latent variables and the complete-data likelihood}

In many statistical models, the observed data \(\bm{X}\) alone do not fully reveal the underlying structure of the system\cite{Dempster1977}. It is often useful, in those cases, to postulate the existence of latent variables \(\bm{Y}\), which are unobserved, implicit variables in the adopted model but not directly accessible from measurements. Together, \(\bm{X}\) and \(\bm{Y}\) constitute the \emph{complete data}, whose joint distribution \(P(\bm{X}, \bm{Y}|\mathcal{I})\) is typically more tractable than the observed-data likelihood \(P(\bm{X}|\mathcal{I})\) alone~\cite{bishop2006pattern}.
The relationship between them is made precise by marginalizing the joint distribution over all possible values of \(\bm{Y}\),
\begin{equation}
\label{eq:marginal_likelihood}
P(\bm{X} | \mathcal{I}) = \int d\bm{Y}\,P(\bm{X}, \bm{Y} | \mathcal{I}).
\end{equation}
However, this integral is in general intractable, which motivates the EM approach. Working with $P(\bm X | \mathcal{I})$ as a marginal distribution over the latent variable \(\bm Y\) requires knowledge of the conditional distributions. To make this explicit, we invoke the product rule of probability theory,
\begin{equation}
\label{eq:bayes_identity}
P(\bm X, \bm Y | \mathcal I) = P(\bm X | \mathcal I) \cdot P(\bm Y | \bm X, \mathcal I),
\end{equation}
which relates the joint distribution to the conditional densities of both the observed and latent variables.


%
%
%
\subsection{Expectation-Maximization Algorithm steps}
\label{sec:EM-Algorithm}

Once we have adopted a joint distribution \(P(\bm X, \bm Y| \bm \lambda)\), the goal is to find the parameter values \(\hat{\bm \lambda_{EM}}\) that maximize the complete-data log-likelihood \(\ln P( \bm X, \bm Y |\bm \lambda)\).
Since \(\bm Y\) is unobserved, this is accomplished indirectly, by iterating through the \emph{E} (expectation) and \emph{M} (maximization) steps of the \emph{EM algorithm}.

\vspace{10pt}
\textbf{E-step:} Let $P(\bm Y | \bm X, \bm \lambda^{(t)})$ be the conditional distribution for the latent variable $\mathbf Y$ given the observed data $\bm X$ and the parameter estimate $\bm{\lambda}^{(t)}$ at the current step $t$. In the E-step we compute the expectation of the complete-data log-likelihood 
$\ln P(\bm X, \bm Y |\bm \lambda)$ as a function of $\bm \lambda$ keeping $\bm{\lambda}^{(t)}$ fixed. That is, we define a new function
\begin{equation}
\label{eq:Q_function}
\begin{split}
Q(\bm \lambda; \bm{\lambda}^{(t)}) := \Big<\ln P(\bm X, \bm Y | \bm\lambda)\Big>_{\bm X, \bm{\lambda}^{(t)}} = \int d\bm Y\,P(\bm Y | \bm X, \bm{\lambda}^{(t)})\ln P(\bm X, \bm Y | \bm \lambda)
\end{split}
\end{equation}
that is used in the next step.

\vspace{10pt}
\textbf{M-step:} This step consists in finding the value of $\bm \lambda$ that maximizes the expected complete-data log-likelihood $Q(\bm \lambda|\bm{\lambda}^{(t)})$ computed in the E-step, and assign it as 
$\bm{\lambda}^{(t+1)}$. In other words, we update $\bm \lambda$ by using the rule
\begin{equation}
\bm{\lambda}^{(t+1)} = \arg\max_{\bm \lambda} \, Q(\bm \lambda; \bm{\lambda}^{(t)}).
\end{equation}

The algorithm is initialized with a parameter value $\bm{\lambda}^{(0)}$ and proceeds by alternating the E- and M-steps until a prescribed convergence criterion is met, and we denote the final estimate after convergence as \(\hat{\bm{\lambda}}_\text{EM}\). Each iteration $t$ produces a new estimate $\bm{\lambda}^{(t+1)}$ in such a way that 

\begin{equation}
\mathcal{L}(\bm{\lambda}^{(t+1)}|\bm X) \geq \mathcal{L}(\bm{\lambda}^{(t)}|\bm X),
\end{equation}
so that the sequence $\bm{\lambda}^{(0)}, \bm{\lambda}^{(1)}, \bm{\lambda}^{(2)}, \ldots$ converges~\cite{Dempster1977} to a stationary point of the observed-data log-likelihood $\ln \mathcal{L}(\bm \lambda|\bm X)$.

The original EM article \cite{Dempster1977} surveys a wide range of applications and modeling techniques, among which Section 4.6 presents the hierarchical Gaussian observation with a gamma prior on the precision as the prototypical instance of EM for iteratively reweighted least squares. Under a direct identification of variables, that construction corresponds to the one-dimensional, single-observation form of the kappa-superstatistics problem treated here. Beyond it, the next section pursues three further steps: the E-step is derived for arbitrary dimensionality \(d\) of the velocity vector accessible to the diagnostic, the M-step is developed explicitly as the joint maximization over the gamma hyperparameters \((\alpha,\theta)\), and the resulting algorithm is embedded in the Beck-Cohen superstatistical framework, where these hyperparameters acquire direct physical meaning as the parameters of the prior on the inverse temperature.

\section{Kappa Parameter Estimation via the EM Algorithm in the Superstatistics Framework}
\label{sec:EM_algorithm_SS}

Having reviewed the general structure of the EM algorithm, we now cast the kappa distribution into this framework. As established in Section~\ref{sec:superstatistics}, the particle velocities \(\bm{v}_i\) are the observable variables, while the inverse temperature \(\beta\) plays the role of the latent (unobserved) variable replacing the generic $\bm Y$ of Section~3.2 by the concrete inverse-temperature vector $\bm B = (\beta_1, \ldots, \beta_N)$. Within the superstatistical framework, each velocity datum 
\(\bm{v}_i\) is paired with its own value of inverse temperature \(\beta_i\). We will consider the pairs \(\{(\bm{v}_i, \beta_i)\}\) as mutually independent, and that the latent distribution 
is governed by the hyperparameter vector \(\bm{\lambda} := (\alpha, \theta)\).

In order to make an explicit representation of \(Q(\alpha, \theta; \alpha^{(t)}, \theta^{(t)})\) in  \eqref{eq:Q_function}, we require the joint distribution \(P(\bm V, \bm B|\alpha, \theta)\) where $\bm{B} = (\beta_1, \beta_2, \ldots, \beta_N)$, as well as the distribution of latent variables \(P(\bm B|\bm X, \alpha, \theta)\). Given that each pair is statistically independent, this factorizes as
\begin{equation}
\label{eq:joint_distribution_kappa_EM_SS}
P(\bm V, \bm B|\alpha, \theta) = \prod_{i=1}^N P(\bm v_i, \beta_i |\alpha, \theta)
\end{equation}
The complete-data distribution is obtained by combining the Maxwell-Boltzmann likelihood~\eqref{eq:maxwell_distribution_of_velocities} with the gamma distribution \eqref{eq:gamma_distribution_of_beta} on the latent variable 
$\beta$,
\begin{equation}
\label{eq:joint_distribution_velocity_beta}
 \begin{split}
  P(\bm{v_i}, \beta_i | \alpha, \theta) 
  &= P(\beta_i | \alpha, \theta)\, P(\bm{v_i} | \beta_i)\\[6pt]
  &= \frac{(m/2\pi)^{\frac{d}{2}}}{\theta^{\alpha}\,\Gamma(\alpha)}\,
     \beta_i^{\alpha + \frac{d}{2}-1}\,
     \exp\!\left[-\beta_i\left(\frac{1}{\theta} + \frac{m}{2}\bm{v_i}^2\right)\right],
 \end{split}
\end{equation}

so that, regarded as a function of \(\beta_i\), this expression is proportional to a gamma distribution; this conjugacy is what renders the E-step analytically tractable.

The distribution of \(\beta\) given the velocities $\bm{V}$ and the current hyperparameters $\bm{\lambda}^{(t)} = (\alpha^{(t)}, \theta^{(t)})$ is obtained by Bayes' theorem, using the factorization of the joint distribution in \eqref{eq:joint_distribution_kappa_EM_SS},
\begin{equation}
\label{eq:posterior_latent_distribution_EM_SS}
P(\boldsymbol{B} | \bm{V}, \bm{\lambda}^{(t)})
= \frac{\prod_i P(\bm{v}_i, \beta_i | \bm{\lambda}^{(t)})}
       {\prod_i P(\bm{v}_i | \bm{\lambda}^{(t)})}
= C \prod_i P(\bm{v}_i, \beta_i | \bm{\lambda}^{(t)}),
\end{equation}
where
\begin{equation}
C = \frac{1}{\prod_{i=1}^N P(\bm{v}_i | \bm{\lambda}^{(t)})}
\end{equation}
collects all factors that depend neither on \(\bm{B}\) nor on the parameters \((\alpha, \theta)\) to be optimized.
Comparing with \eqref{eq:joint_distribution_velocity_beta}, each individual factor \(P(\bm{v}_i, \beta_i | \lambda^{(t)})\) regarded as a function of \(\beta_i\), is proportional to a gamma
distribution with updated parameters
\begin{align}
\label{eq:alpha_g_theta_g}
\alphag & = \alpha^{(t)} + \frac{d}{2}, \\
\thetag & = \left(\frac{1}{\theta^{(t)}} + \frac{m}{2}\,\bm{v}_i^2\right)^{-1},
\end{align}
where \(d\) is the dimensionality of the velocity data accessible to the diagnostic; the implementation reported in Sec.~\ref{sec:results} adopts \(d=1\), corresponding to a single-component velocity diagnostic. So this distribution factorizes as \(P(\bm{B} | \bm{V}, \bm{\lambda}^{(t)}) = \prod_i P(\beta_i | \bm v_i, \alphag, \thetag)\).


\subsection{E-step: construction of \(Q(\alpha, \theta ; \alpha^{(t)}, \theta^{(t)})\)}

Replacing the specific representations in \eqref{eq:joint_distribution_kappa_EM_SS} and \eqref{eq:posterior_latent_distribution_EM_SS} into the function $Q$ in \eqref{eq:Q_function} and, 
writing \(\bm \lambda \defeq (\alpha, \theta)\), \(\bm{\lambda}^{(t)} \defeq (\alpha^{(t)}, \theta^{(t)})\), we have
\begin{equation*}
 \begin{split}
  Q(\bm \lambda ; \bm{\lambda}^{(t)}) & = C\int d\bm{B}\,P(\bm{V}, \bm{B} | \bm{\lambda}^{(t)})\ln P(\bm{V}, \bm{B} | \bm \lambda) \\
  & = C \int \prod_{j=1}^N d\beta_j\,P(\bm{v}_j, \beta_j | \bm{\lambda}^{(t)})\left\{\sum_{i=1}^N \ln P(\bm{v}_i, \beta_i | \bm \lambda)\right\} \\
  & = C \sum_i \int d\beta_i\,P(\bm{v}_i, \beta_i | \bm{\lambda}^{(t)})\ln P(\bm{v}_i, \beta_i | \bm \lambda)\,
      \left[ \prod_{j \neq i} \int d\beta_j\, P(\bm{v}_j, \beta_j | \bm{\lambda}^{(t)})\right].
 \end{split}
\end{equation*}
 
Each factor in the expression in brackets on the last line integrates to \(P(\bm{v}_j | \bm{\lambda}^{(t)})\), so the product \(\prod_{j\neq i} P(\bm{v}_j | \bm{\lambda}^{(t)})\) is absorbed into 
the constant \(C\). Since this prefactor depends neither on \(\beta_i\) nor on the parameters \(\bm \lambda\) being optimized, it does not affect the maximization and can be dropped, yielding
\begin{equation}
\label{eq:Q_decoupled}
Q(\alpha, \theta ; \alpha^{(t)}, \theta^{(t)}) = C \sum_{i=1}^N \int d\beta_i\; P(\bm{v}_i, \beta_i | \alpha^{(t)}, \theta^{(t)})\ln P(\bm{v}_i, \beta_i | \alpha, \theta).
\end{equation}

Substituting the explicit form of the joint distribution \eqref{eq:joint_distribution_velocity_beta} and  \eqref{eq:posterior_latent_distribution_EM_SS} into~\eqref{eq:Q_decoupled} and expanding the logarithm, terms independent of $\beta_i$ can be factored outside the integral directly, while the remaining terms reduce to expectations with respect to $\mathrm{Gamma}(\beta_i \mid \alphag, \thetag)$, namely
\begin{subequations}
\begin{align}
\big<\beta_i\big>_{\alphag,\thetag} & = \alphag\,\thetag, \\
\big<\ln\,\beta_i \big>_{\alphag,\thetag} & = \psi(\alphag) + \ln\thetag,
\end{align}
\end{subequations}
where $\psi$ denotes the digamma function. Collecting terms finally yields
\begin{equation}
	\label{eq:Q_kappa}
	\begin{split}
		Q(\alpha,\theta ; \alpha^{(t)}, \theta^{(t)}) = \sum_{i=1}^N \Bigg\{\frac{d}{2}\ln\!\left(\frac{m}{2\pi}\right)
		+ & \left(\alpha+\tfrac{d-2}{2}\right) \Big[\psi(\alphag)+\ln\thetag\Big] \\
		& - \alpha\ln\theta - \ln\Gamma(\alpha) - \alphag\,\thetag \left(\frac{1}{\theta} + \frac{m}{2}\,\bm{v}_i^2\right)\Bigg\}.
	\end{split}
\end{equation}
A direct computation shows that both extremum conditions \eqref{eq:update_theta} and \eqref{eq:update_alpha} are independent of \(d\). In \(\partial Q / \partial \theta\) the dimensionality term \((d/2)\ln(m/2\pi)\) is constant in \(\theta\) and the coefficient \(\alpha + (d-2)/2\) does not contain \(\theta\), whereas in \(\partial Q / \partial \alpha\) the same coefficient differentiates to unity, eliminating any explicit dependence on \(d\). 
The dimensionality therefore enters the algorithm only through the posterior shape \(\tilde\alpha = \alpha^{(t)} + d/2\) and through the final conversion \(\hat\kappa = \hat\alpha + (d-2)/2\), as anticipated in Section~\ref{sec:superstatistics}.

%
%
%
%
\subsection{M-step: maximization of $Q$ with respect to $\alpha$ and $\theta$}

In the M-step, $Q(\alpha,\theta ; \alpha^{(t)}, \theta^{(t)})$ is maximized with respect to $\alpha$ and $\theta$. Note that the parameters $\alphag$ and $\thetag$ depend on \(\bm{v}_i\) and 
$(\alpha^{(t)}, \theta^{(t)})$, as we can see in the~\eqref{eq:alpha_g_theta_g}, so they are fixed quantities during this step.

\noindent
The equations for the extremum of $Q(\alpha, \theta|\alpha^{(t)}, \theta^{(t)})$ are
\begin{equation}
\frac{\partial}{\partial \theta}Q(\alpha, \theta;\alpha^{(t)}, \theta^{(t)}) = \sum_{i=1}^N \left(\frac{\alphag\,\thetag}{\theta^2} - \frac{\alpha}{\theta}\right) = 0,
\end{equation}
which yields the closed-form update
\begin{equation}
\label{eq:update_theta}
\theta_\text{new} = \frac{\alphag}{\alpha_\text{new}\,N}\sum_{i=1}^N \thetag,
\end{equation}
%
and, similarly,
\begin{equation}
\frac{\partial}{\partial \alpha}Q(\alpha, \theta|\alpha^{(t)}, \theta^{(t)}) = \sum_{i=1}^N \Bigl[\psi(\alphag)+\ln\thetag -\ln\theta - \psi(\alpha)\Bigr] = 0,
\end{equation}
which, upon substituting the expression for $\theta_\text{new}$ from~\eqref{eq:update_theta} and rearranging, leads to
\begin{equation}
\label{eq:update_alpha}
\psi(\alpha) - \ln\alpha = \frac{1}{N}\sum_{i=1}^N \Big[\psi(\alphag)+\ln\thetag\Big] - \ln\!\left(\frac{\alphag}{N}\sum_{i=1}^N \thetag\right).
\end{equation}

The right-hand side is a constant that can actually be computed from the current data and $(\alpha^{(t)}, \theta^{(t)})$. Since $\psi(\alpha)-\ln\alpha$ is a monotonically decreasing 
function of $\alpha$, it follows that \eqref{eq:update_alpha} has a unique solution, which must be found numerically. Once $\alpha^{(t+1)}$ is obtained, $\theta^{(t+1)}$ follows 
immediately from \eqref{eq:update_theta}.

Finally, optimal estimation of the target parameter \(\kappa\), that we will denote by $\hat{\kappa}$, is recovered from the converged estimate \(\hat{\alpha}\) via 
\eqref{eq:relation_kappa_alpha_theta},
\begin{equation*}
\hat \kappa = \hat \alpha +\frac{(d-2)}{2},
\end{equation*}
completing the full loop of the EM algorithm.


%
%
%
\subsection{Parameter Initialization}
\label{sec:Parameter_initialization}

Improved initial estimates \(\alpha_0\) and \(\theta_0\) can be obtained from the empirical moments of the velocity data. Within the superstatistical framework, the moments of the single-particle kinetic energy distribution under a gamma-distributed inverse temperature have been derived in closed form~\cite{Davis2026}. Expressed in terms of the gamma parameters \((\alpha, \theta)\) via the parameterization in \eqref{eq:kappa_marginal_alpha_theta}, 
the second and fourth moments of a single velocity component are given by
\begin{subequations}
\begin{align}
\label{eq:v2_1D}
\big<v_x^2\big>_{\alpha, \theta} & = \frac{1}{m\theta(\alpha - 1)},  \qquad \alpha > 1, \\
\label{eq:v4_1D}
\big<v_x^4\big>_{\alpha, \theta} & = \frac{3}{m^2\theta^2(\alpha-1)(\alpha-2)},  \qquad \alpha > 2 \quad (\kappa > 5/2),
\end{align}
\end{subequations}
while for the three-dimensional speed \(v = |\bm{v}|\) we have
\begin{align}
\label{eq:v2v4_3D}
\big<v^2\big>_{\alpha, \theta} & = \frac{3}{m\theta(\alpha - 1)}, \\
\big<v^4\big>_{\alpha, \theta} & = \frac{15}{m^2\theta^2(\alpha-1)(\alpha-2)}.
\end{align}

\noindent
In both cases the scale parameter \(\theta\) cancels in the dimensionless moment ratio
\begin{equation}
\label{eq:moment_ratio}
R \;\defeq\; \frac{\big<v^4\big>_{\alpha, \theta}}{\big<v^2\big>_{\alpha, \theta}^2} \; = \; \frac{(d+2)(\alpha - 1)}{d\,(\alpha - 2)},
\end{equation}
where \(d\) is the number of velocity components used (\(d = 1\) for a single component, \(d = 3\) for the speed). Inverting for \(\alpha\) gives
\begin{equation}
\label{eq:alpha_init}
\alpha_0 =
\begin{cases}
  \displaystyle\frac{2R - 3}{R - 3}, & d = 1, \\[12pt]
  \displaystyle\frac{6R - 5}{3R - 5}, & d = 3,
\end{cases}
\end{equation}
which is valid for \(R > (d+2)/d\); the lower bound corresponds to the Maxwellian limit \(\alpha \to \infty\). Once \(\alpha_0\) is fixed, the initial scale parameter $\theta_0$ follows from the mean kinetic energy 
per component,
\begin{equation}
\label{eq:theta_init}
\theta_0 = \frac{d}{2\,\langle E_k \rangle\,(\alpha_0 - 1)},
\end{equation}
where \(\big<E_k\big> = \frac{m}{2N}\sum_{i=1}^{N} v_i^2\).

If the empirical ratio satisfies \(R \le (d+2)/d\), or the resulting \(\alpha_0 \leq 2\) (equivalently \(\kappa \leq (d+2)/2\)), that is \(\kappa\leq 3/2\) for \(d=1\) and \(\kappa \leq 5/2\) for \(d = 3\), the moment-based estimate is unreliable. This could happen either because the data are nearly Maxwellian, or because \(\kappa\) is too low for the fourth moment to exist (\(\alpha > 2\) is required). To stabilize the initialization in this regime, our implementation switches to an alternative estimator based on the logarithmic moments of the kinetic energy, which remain finite for all \(\alpha > 0\); the two estimators are combined into a hybrid routine \textsc{InitHybrid} that selects the most reliable branch depending on the empirical value of \(R\), and is invoked as a single call in Algorithm~\ref{alg:em-kappa}.

The moment-based initializer admits a direct interpretation within the \((u, \beta_S)\) parameterization introduced in \cite{Davis2026}. Since \(u = 1/\alpha\) is the relative variance of the inverse temperature distribution, and the moment ratio \(R\) yields \(\alpha_0\) via~(\ref{eq:alpha_init}), the initializer effectively estimates the magnitude of the fluctuations of \(\beta\) directly from the 
velocity data, as
\begin{equation}
\label{eq:u_from_alpha}
u_0 = \frac{1}{\alpha_0}.
\end{equation}
In one dimension this gives explicitly \(u_0 = (R - 3)/(2R - 3)\). This relation reveals a conceptual continuity between the initialization and the iterative refinement performed by the EM algorithm: the moment ratio provides a first estimate of the relative variance \(u\) of the latent inverse temperature, which the algorithm subsequently refines through the distributions constructed at each iteration. In the Maxwellian limit (\(\alpha \to \infty\)) we have \(R \to (d+2)/d\) and \(u_0 \to 0\), reflecting the absence of temperature fluctuations; conversely, small \(\alpha\) produces large \(R\) and \(u_0 \to 1\), signaling strong heterogeneity in the inverse temperature. The moment ratio thus serves as a first diagnostic of the degree of non-equilibrium in the system, accessible without any iterative computation.

A final clarification concerns the scope of the gamma assumption on $P(\beta\mid\alpha,\theta)$. The framework is agnostic about the physical mechanism generating fluctuations of $\beta$, but it does assume a gamma functional form for the prior; this choice is what yields the kappa marginal, so it is a modeling assumption tied to the target distribution rather than an empirical claim about $\beta$ itself. When applied to data for which the underlying prior is not gamma, the estimator returns the maximum-likelihood kappa fit to the empirical distribution, in the same spirit as the heuristic kappa fits routinely performed on plasma data \cite{Binsack1966, Vasyliunas1968}, but with explicit probabilistic structure and the consistency guarantees of a properly derived MLE. For systems that are superstatistical by construction, such as numerical simulations with a stochastically driven inverse temperature or models within the Beck-Cohen framework, the gamma assumption holds exactly and the estimator recovers a genuine physical parameter rather than a fitted summary statistic.
%
%
\subsection{Numerical implementation of the superstatistical EM algorithm}

The superstatistical EM algorithm is implemented following the steps of Section~\ref{sec:EM_algorithm_SS}: at each iteration the posterior parameters of the latent variables are computed and the averages \(S_1\) and \(S_2\) are constructed; \(\alpha\) is updated by numerically solving the monotonic equation~\eqref{eq:update_alpha}, \(\theta\) follows from~\eqref{eq:update_theta}, and the iteration stops when the relative change of both parameters falls below a tolerance \(\varepsilon_{\text{param}}\). The complete pseudocode is presented in Algorithm~\ref{alg:em-kappa} for the one-dimensional case (\(d = 1\)), in which the input is a sample of single-component velocities and the posterior shape reduces to \(\tilde\alpha = \alpha + 1/2\); the extension to higher-dimensional diagnostics is straightforward and requires only adapting the input format and using \(\tilde\alpha = \alpha + d/2\) in the E-step, while the M-step update equations remain unchanged.

\begin{algorithm}[t!]
\caption{EM Algorithm for Kappa Distribution Parameter 1D Estimation.}
\label{alg:em-kappa}
\small
\begin{algorithmic}[1]

\Require Velocities $\mathbf{v}=\{v_i\}_{i=1}^N$, mass $m$, optional initial $(\alpha_0,\theta_0)$, tolerance $\varepsilon_{\text{param}}$, max iterations $I_{\max}$
\Ensure Estimated parameters $\alpha$, $\theta$, and $\kappa = \alpha - \frac12$

\State \textbf{Initialization}
\If{$\alpha_0,\theta_0$ not given}
    \State $(\alpha, \theta) \gets \textsc{InitHybrid}(\mathbf{v}, m)$  \Comment{see Sec.~\ref{sec:Parameter_initialization}}
\Else
    \State $\alpha \gets \alpha_0$, $\theta \gets \theta_0$
\EndIf

\For{$iter = 1$ \textbf{to} $I_{\max}$}
    \State \textbf{E-step:} Expectation over $\beta$
    \For{$i = 1$ \textbf{to} $N$}
        \State $\alphag[i] \gets \alpha + 0.5$               \Comment{for $d =1$: $\alphag = \alpha +d/2 = \alpha +1/2$}
        \State $\thetag[i] \gets \dfrac{1}{1/\theta + m v_i^2/2}$
    \EndFor

    \State $S_1 \gets \frac{\alphag}{N}\sum_{i=1}^N \thetag[i]$
    \State $S_2 \gets \frac{1}{N}\sum_{i=1}^N \bigl[\psi(\alphag[i]) + \ln\thetag[i]\bigr]$  \Comment{$\psi$ is digamma}

    \State \textbf{M-step:} Maximization of $Q$
    \State Solve $\psi(\alpha_{\text{new}}) - \ln\alpha_{\text{new}} = S_2 - \ln S_1$ for $\alpha_{\text{new}} > 1$ \Comment{use any numeric-method}
    \State $\theta_{\text{new}} \gets S_1 / \alpha_{\text{new}}$

    \State \textbf{Convergence check}
    \State $\varepsilon \gets 10^{-15}$
    \State $\delta_\alpha \gets \dfrac{|\alpha_{\text{new}} - \alpha|}{|\alpha| + \varepsilon}$
    \State $\delta_\theta \gets \dfrac{|\theta_{\text{new}} - \theta|}{|\theta| + \varepsilon}$
    \If{$\delta_\alpha < \varepsilon_{\text{param}}$ \textbf{and} $\delta_\theta < \varepsilon_{\text{param}}$}
        \State \textbf{break}
    \EndIf

    \State $\alpha \gets \alpha_{\text{new}}$
    \State $\theta \gets \theta_{\text{new}}$
\EndFor

\State $\kappa \gets \alpha - 0.5$
\State \Return $\alpha$, $\theta$, $\kappa$

\end{algorithmic}
\end{algorithm}


\section{Results}
\label{sec:results}


\subsection{Synthetic data generation}
\label{sec:datageneration}

To validate the EM estimator on data of known provenance, we generated synthetic samples \(\bm{V} = \{v_1, v_2, \ldots, v_N\}\), where \(v_i\) is a one-component velocity. This samples were generated directly from the same hierarchical model that underlies the algorithm. The factorization adopted in ~(\ref{eq:joint_distribution_kappa_EM_SS}),
\begin{equation*}
	P(\bm{V}, \bm{B} \,|\, \alpha, \theta) \;=\; \prod_{i=1}^{N} 
	P(v_i, \beta_i \,|\, \alpha, \theta),
\end{equation*}
is the structural assumption that pairs each particle with its own latent inverse temperature \(\beta_i\), drawn independently from the gamma prior, and that justifies treating the data likelihood as a product over single-particle factors. 
Under this model the velocities are independent and identically distributed according to the kappa marginal of~(\ref{eq:marginal_super_kappa_gamma}), and a faithful synthetic dataset is obtained by simulating the same hierarchy in the forward direction:
\begin{equation}
	\beta_i \;\sim\; \mathrm{Gamma}(\alpha,\, \theta), \qquad 
	v_i \,|\, \beta_i \;\sim\; \mathcal{N}\!\left(0,\, \tfrac{1}{m\beta_i}\right).
	\label{eq:forward-sampling}
\end{equation}
This procedure yields exact i.i.d.\ draws from the kappa distribution without recourse to Markov chain methods, and places the validation experiment squarely within the probabilistic model the estimator is designed for.

We adopted a system of natural units with \(m = 1\) and \(K = 0.1\), so that the only physical input determining each dataset is the spectral index \(\kappa\), through the one-dimensional reparameterization \(\alpha = \kappa + 1/2\) and \(\theta = 1/[2 K(\kappa - 1/2)]\). Three target datasets were generated for \(\kappa \in \{2.5,\, 6,\, 12\}\), each consisting of \(N_\text{tot} = 10^{8}\) velocities sampled via ~(\ref{eq:forward-sampling}) 
with a fixed pseudo-random seed; subsamples of size 
\(N \in \{10^{4},\, 10^{5},\, 10^{6}\}\) were drawn from each pool for the statistical study reported in Section~\ref{sec:error-vs-N}.

\subsection{Numerical implementation and validation protocol}
\label{sec:protocol}

The superstatistical EM algorithm of Section~\ref{sec:EM_algorithm_SS} was implemented in Python~3.13 using the NumPy and SciPy scientific computing libraries. The scalar root-finding step in~(\ref{eq:update_alpha}) was 
solved with Brent's method, and the digamma function was evaluated via \texttt{scipy.special.digamma}. Convergence was declared when the relative change of both \(\alpha\) and \(\theta\) fell below \(\varepsilon = 10^{-8}\). The algorithm was initialized with the moment-based scheme of Section~\ref{sec:Parameter_initialization}, with the fallback 
\(\kappa_0 = 6\) used when the empirical fourth-to-second moment ratio failed the physical admissibility condition. All simulations were performed on a workstation with an Intel Core i7-1260P processor and 16~GB of RAM, running Debian GNU/Linux~13.

For each value of \(\kappa\), a Monte Carlo study~\cite{robert2004monte} was performed to characterize the sampling distribution of \(\hat{\kappa}\). From the corresponding pool of \(N_\text{tot} = 10^{8}\) velocities described in Section~\ref{sec:datageneration}, 100 independent subsamples of size \(N\) were drawn without replacement, at each of the three sample sizes 
\(N \in \{10^{4},\, 10^{5},\, 10^{6}\}\). Since \(N \ll N_\text{tot}\), the subsamples are effectively independent and the resulting collection of estimates \(\{\hat{\kappa}^{(j)}\}_{j=1}^{100}\) approximates samples from the true sampling distribution of the EM estimator at sample size \(N\). The algorithm was performed on each subsample, and the reported statistics: mean and standard deviation of \(\hat{\kappa}\), mean iteration count, and mean wall-clock time, were computed over these 100 replicas.

\subsection{Single-run convergence behavior}
\label{sec:single-run}

\begin{figure}[H]
	\centering
	\includegraphics[width=0.8\textwidth]{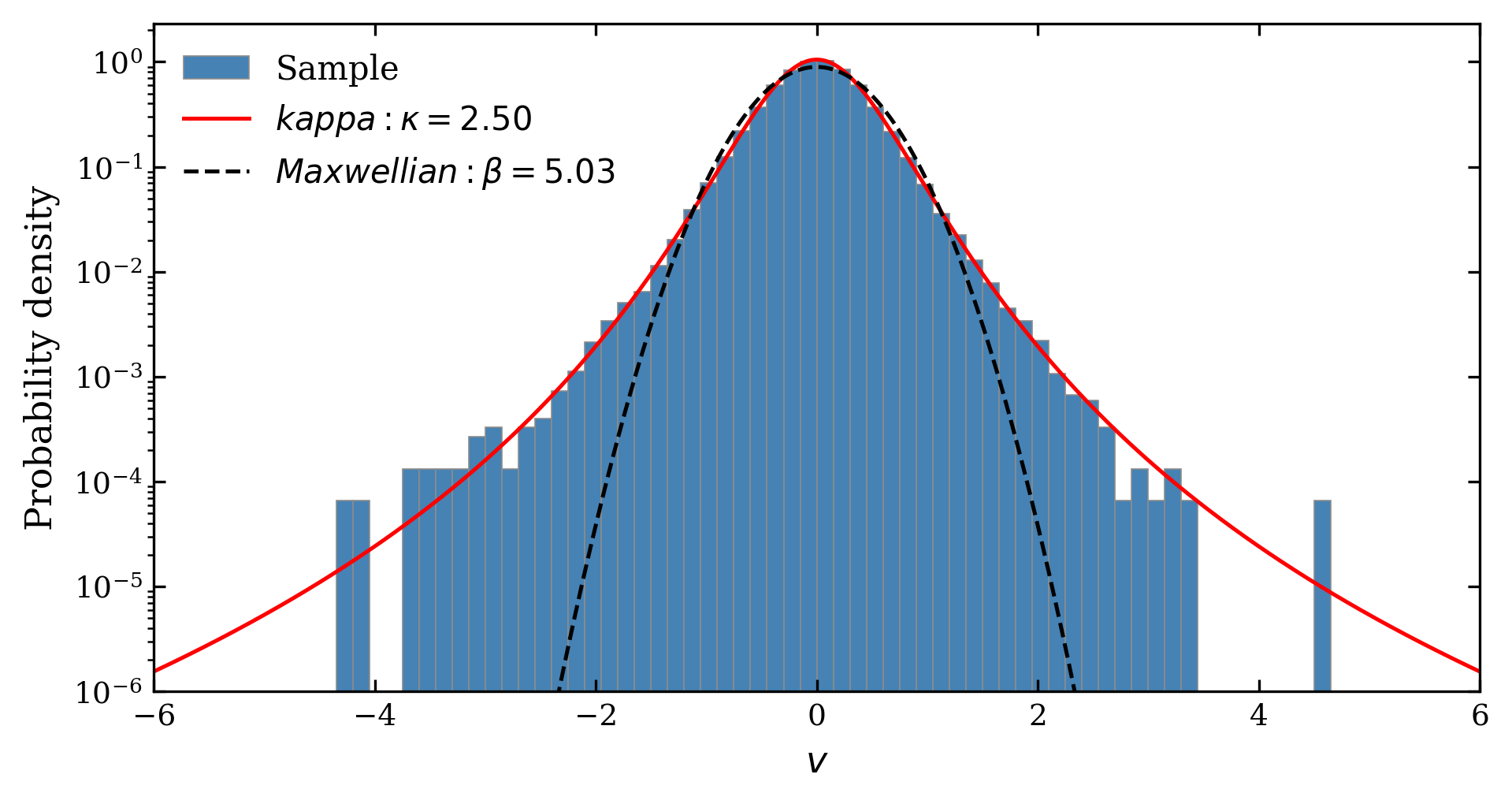}
	\caption{Histogram and optimal kappa distribution of one component of 
		velocity for \(\kappa = 2.5\). The solid red line shows the estimated 
		distribution, with \(\hat{\kappa} = 2.50\).}
	\label{fig:hist-k2.5}
\end{figure}

\begin{figure}[H]
	\centering
	\includegraphics[width=0.8\textwidth]{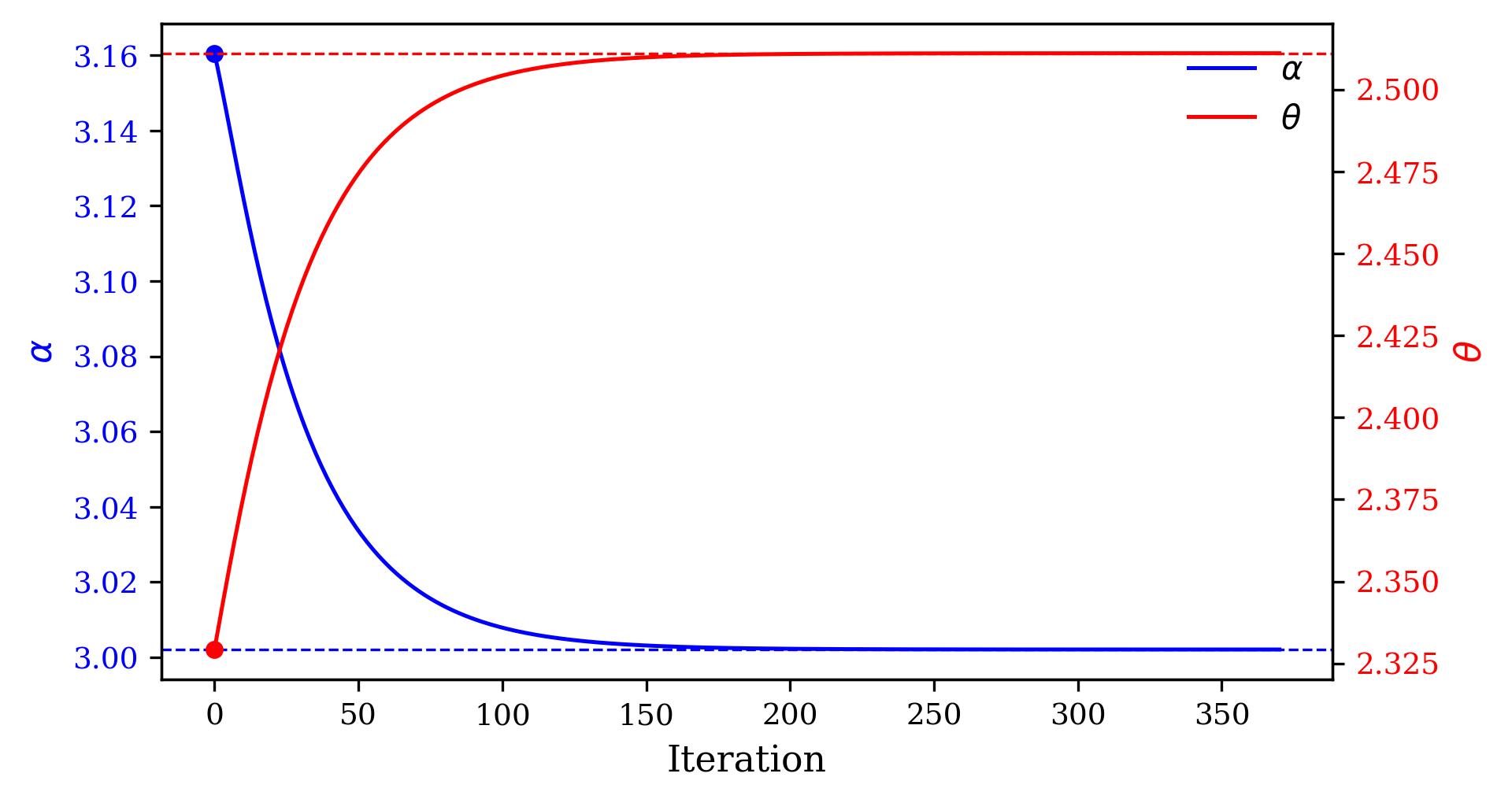}
	\caption{Convergence of \(\alpha\) (left axis, blue) and \(\theta\) 
		(right axis, red) for \(\kappa = 2.5\); true values \(\alpha = 3.0\) and 
		\(\theta = 2.50\). Dots indicate initial values, dashed lines the final 
		estimates.}
	\label{fig:conv-params-k2.5}
\end{figure}

\begin{figure}[H]
	\centering
	\includegraphics[width=0.8\textwidth]{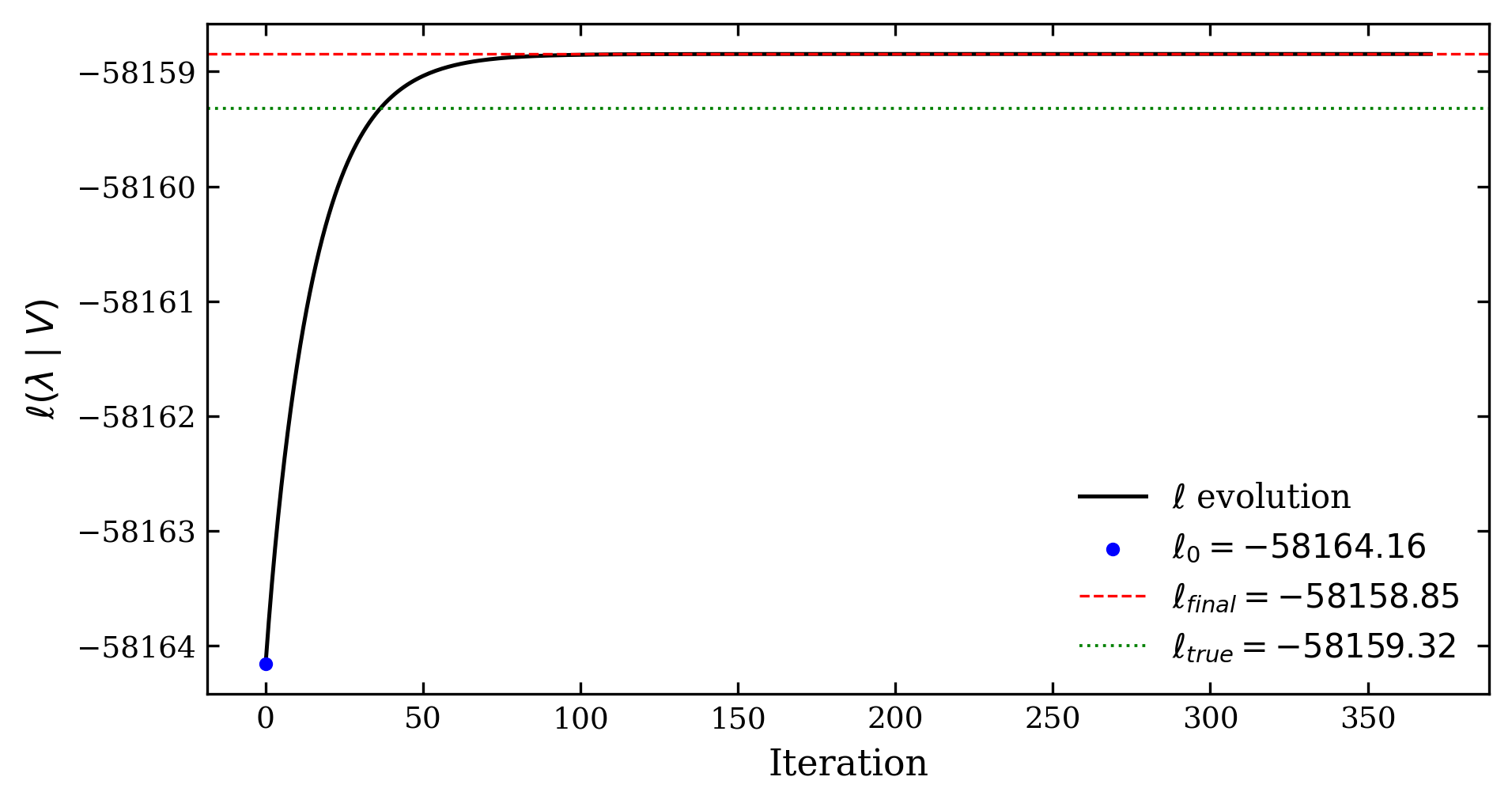}
	\caption{Monotonic increase of the log-likelihood during EM iterations for $\kappa = 2.5$. The dotted green line marks the log-likelihood evaluated at the generating parameters.}
	\label{fig:loglik-k2.5}
\end{figure}

We first examine the qualitative behavior of the EM estimator on a single representative subsample for each of the three values of the spectral index: \(\kappa = 2.5\) (heavy tails), \(\kappa = 6.0\) (intermediate), and 
\(\kappa = 12.0\) (near-Maxwellian). For each case, Figures~\ref{fig:hist-k2.5},~\ref{fig:conv-params-k2.5} 
and~\ref{fig:loglik-k2.5} (and analogously for \(\kappa = 6.0\) and \(12.0\)) show, respectively, the velocity histogram together with the optimal kappa distribution, the convergence of the parameters \(\alpha\) and \(\theta\), and 
the monotonic increase of the log-likelihood during the EM iterations. The red dots mark the moment-based initial values, and the dashed lines indicate the final estimates. All results in this subsection correspond to a single subsample of size \(N = 10^{5}\).

\begin{figure}[H]
	\centering
	\includegraphics[width=0.8\textwidth]{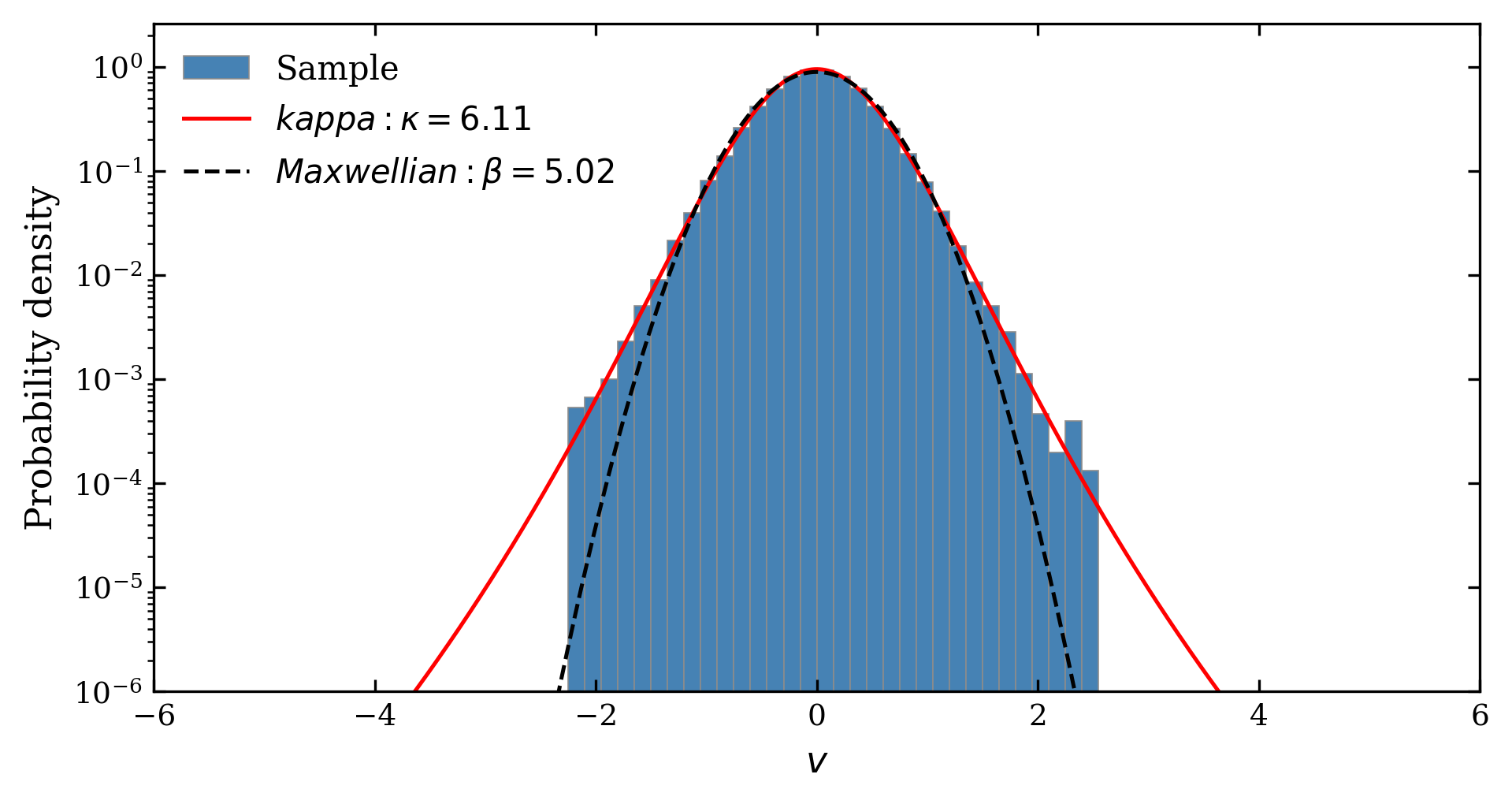}
	\caption{Histogram and optimal kappa distribution for \(\kappa = 6.0\). The 
		solid red line shows the estimated distribution, with 
		\(\hat{\kappa} = 6.11\).}
	\label{fig:hist-k6.0}
\end{figure}

\begin{figure}[H]
	\centering
	\includegraphics[width=0.8\textwidth]{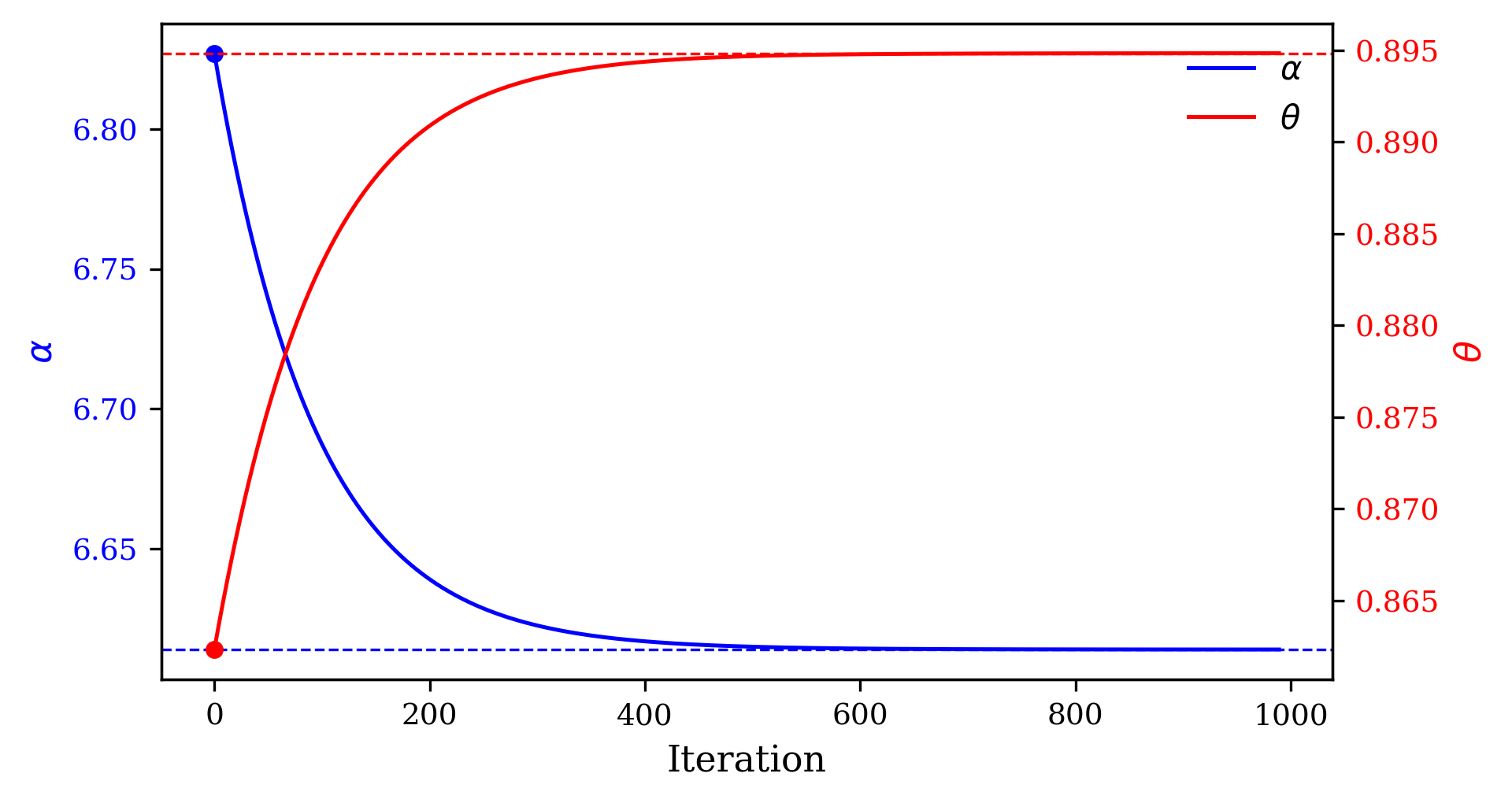}
	\caption{Convergence of \(\alpha\) (left axis, blue) and \(\theta\) 
		(right axis, red) for \(\kappa = 6.0\); true values \(\alpha = 6.5\) and 
		\(\theta = 0.909\). Dots indicate initial values, dashed lines the final 
		estimates.}
	\label{fig:conv-params-k6.0}
\end{figure}

\begin{figure}[H]
	\centering
	\includegraphics[width=0.8\textwidth]{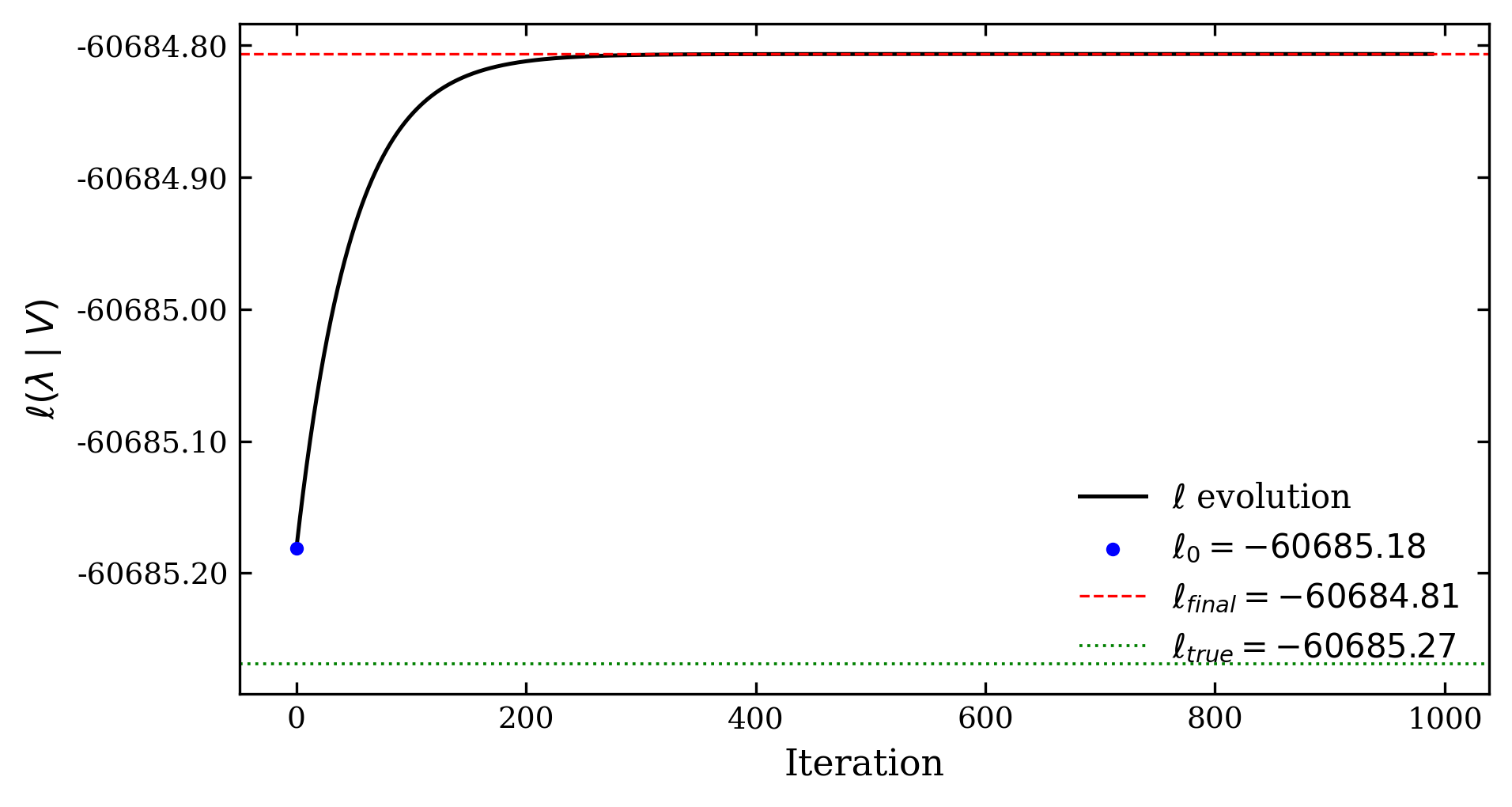}
	\caption{Monotonic increase of the log-likelihood during EM iterations for $\kappa = 6.0$. The dotted green line marks the log-likelihood evaluated at the generating parameters.}
	\label{fig:loglik-k6.0}
\end{figure}

\begin{figure}[H]
	\centering
	\includegraphics[width=0.8\textwidth]{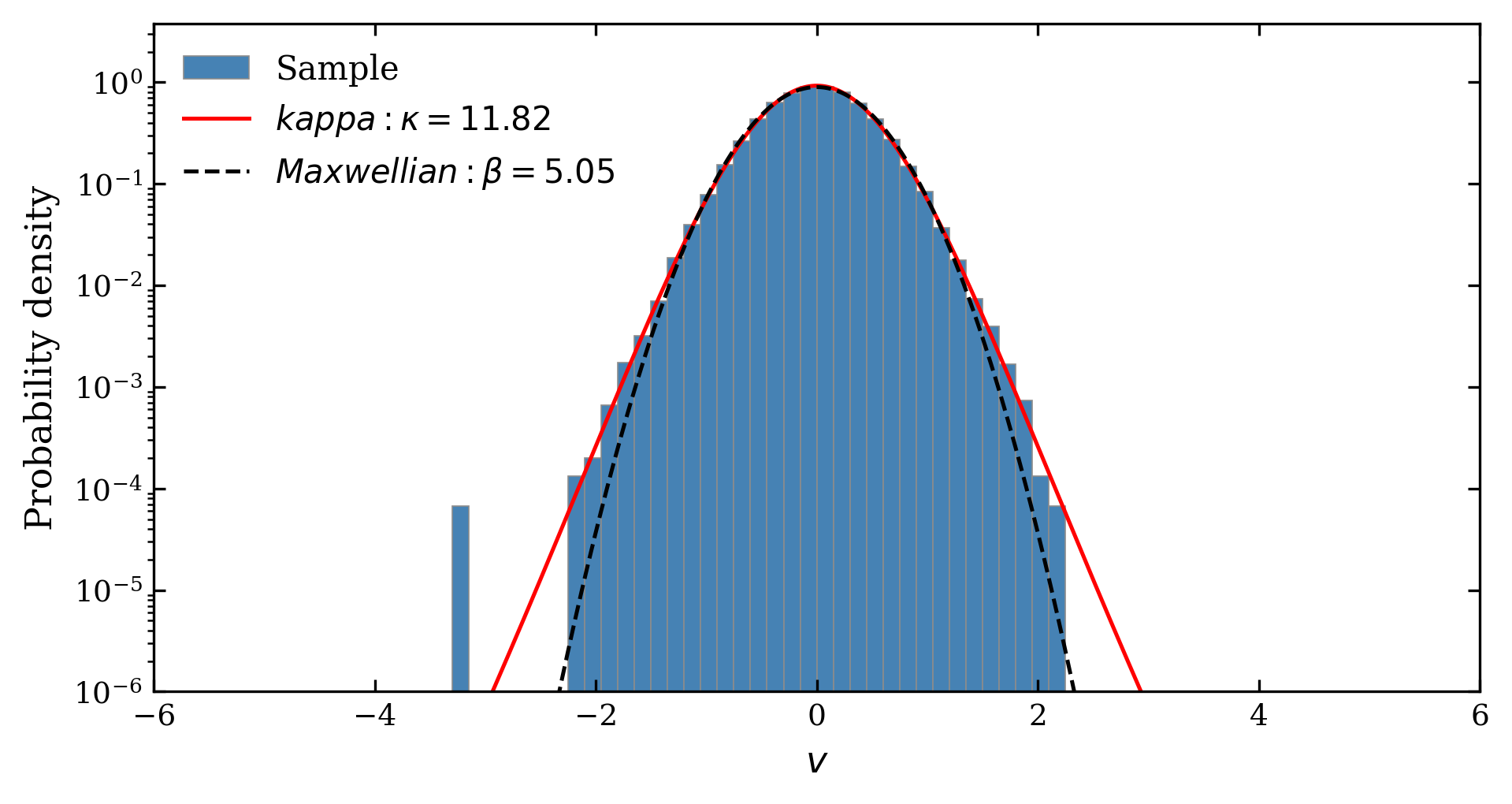}
	\caption{Histogram and optimal kappa distribution for \(\kappa = 12.0\). The 
		solid red line shows the estimated distribution, with 
		\(\hat{\kappa} = 11.82\).}
	\label{fig:hist-k12.0}
\end{figure}

\begin{figure}[H]
	\centering
	\includegraphics[width=0.8\textwidth]{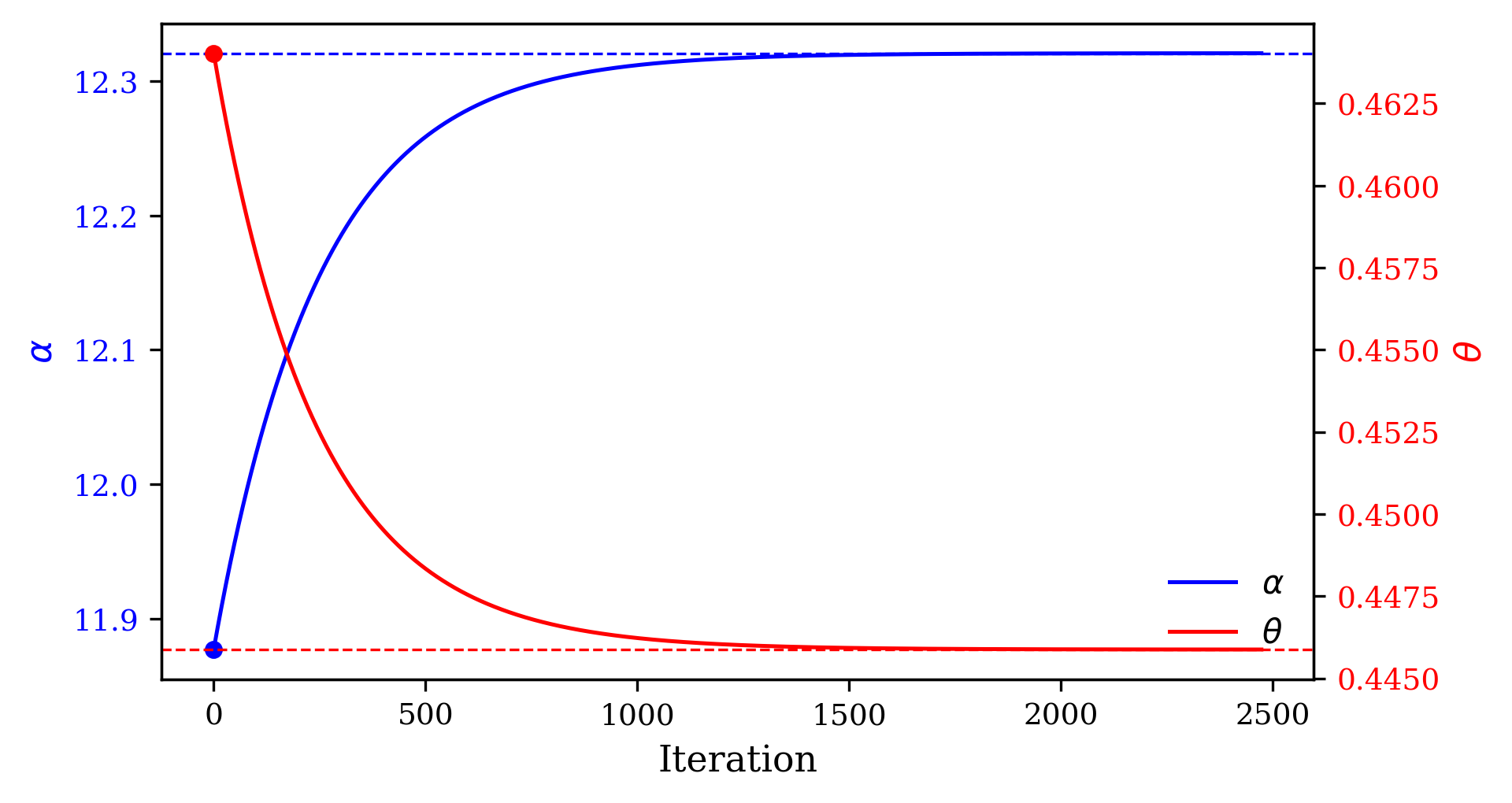}
	\caption{Convergence of \(\alpha\) (left axis, blue) and \(\theta\) 
		(right axis, red) for \(\kappa = 12.0\); true values \(\alpha = 12.5\) and 
		\(\theta = 0.435\). Dots indicate initial values, dashed lines the final 
		estimates.}
	\label{fig:conv-params-k12.0}
\end{figure}

\begin{figure}[H]
	\centering
	\includegraphics[width=0.8\textwidth]{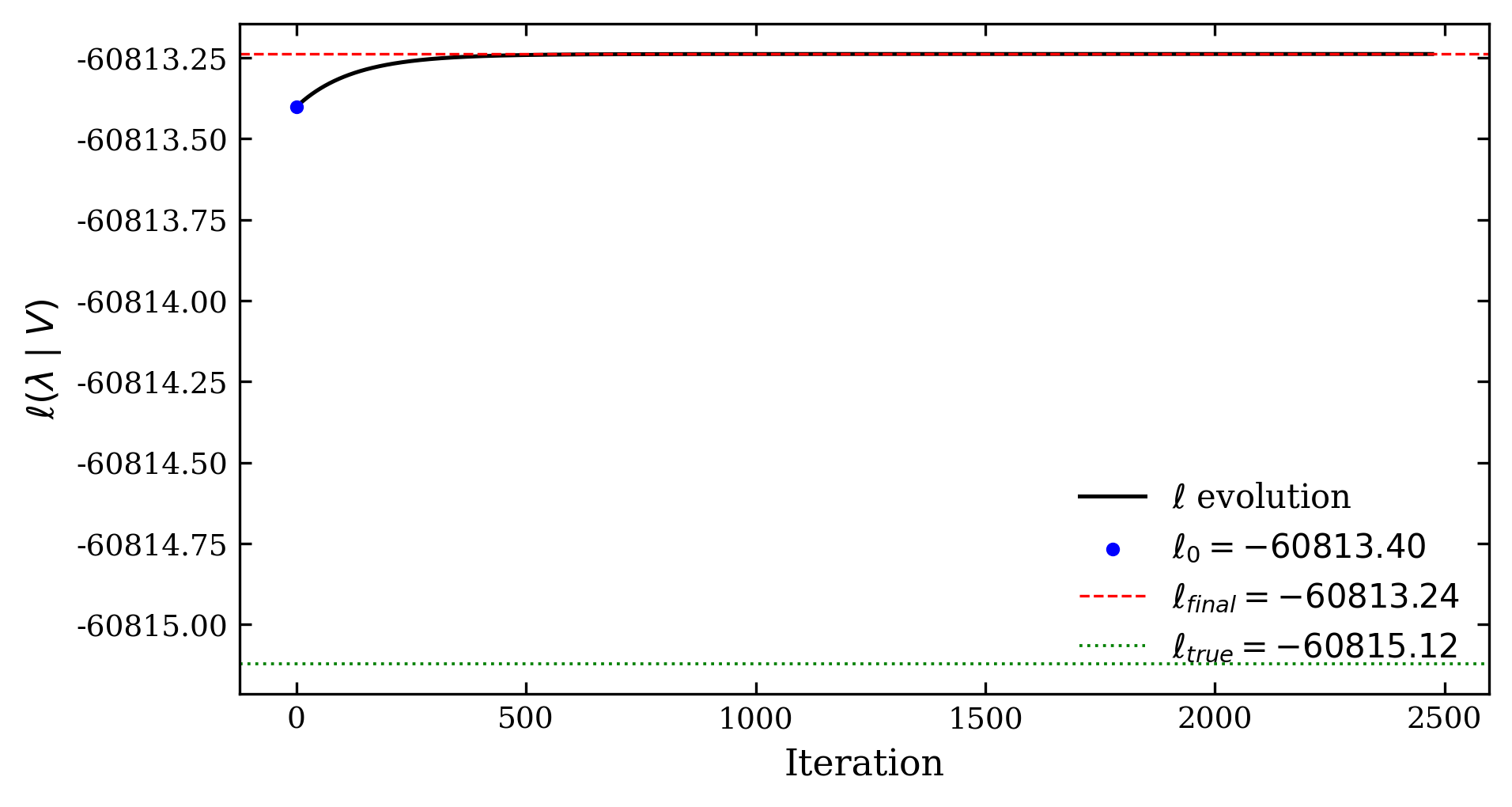}
	\caption{Monotonic increase of the log-likelihood during EM iterations for $\kappa = 12.0$. The dotted green line marks the log-likelihood evaluated at the generating parameters.}
	\label{fig:loglik-k12.0}
\end{figure}

\subsection{Estimation error vs data size}
\label{sec:error-vs-N}

\begin{figure}[H]
	\centering
	\includegraphics[width=0.6\textwidth]{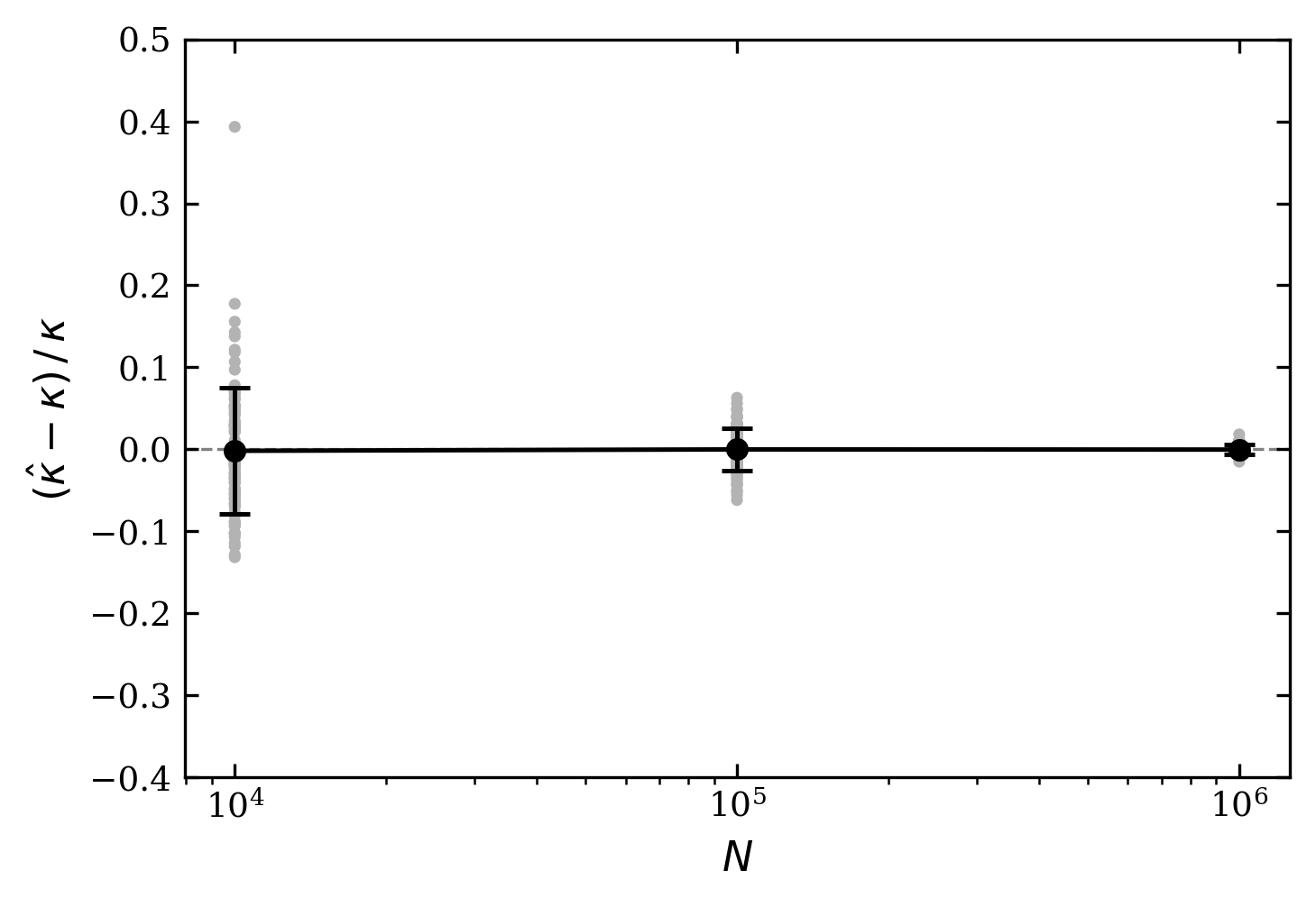}
	\caption{Relative error \((\hat{\kappa} - \kappa)/\kappa\) of the EM estimator 
		as a function of the sample size \(N\), for synthetic data drawn from a kappa 
		distribution with \(\kappa = 2.5\). Each point corresponds to the mean over the 
		synthetic data generated; error bars represent one standard deviation.}
	\label{fig:error-k2.5}
\end{figure}

\begin{figure}[H]
	\centering
	\includegraphics[width=0.6\textwidth]{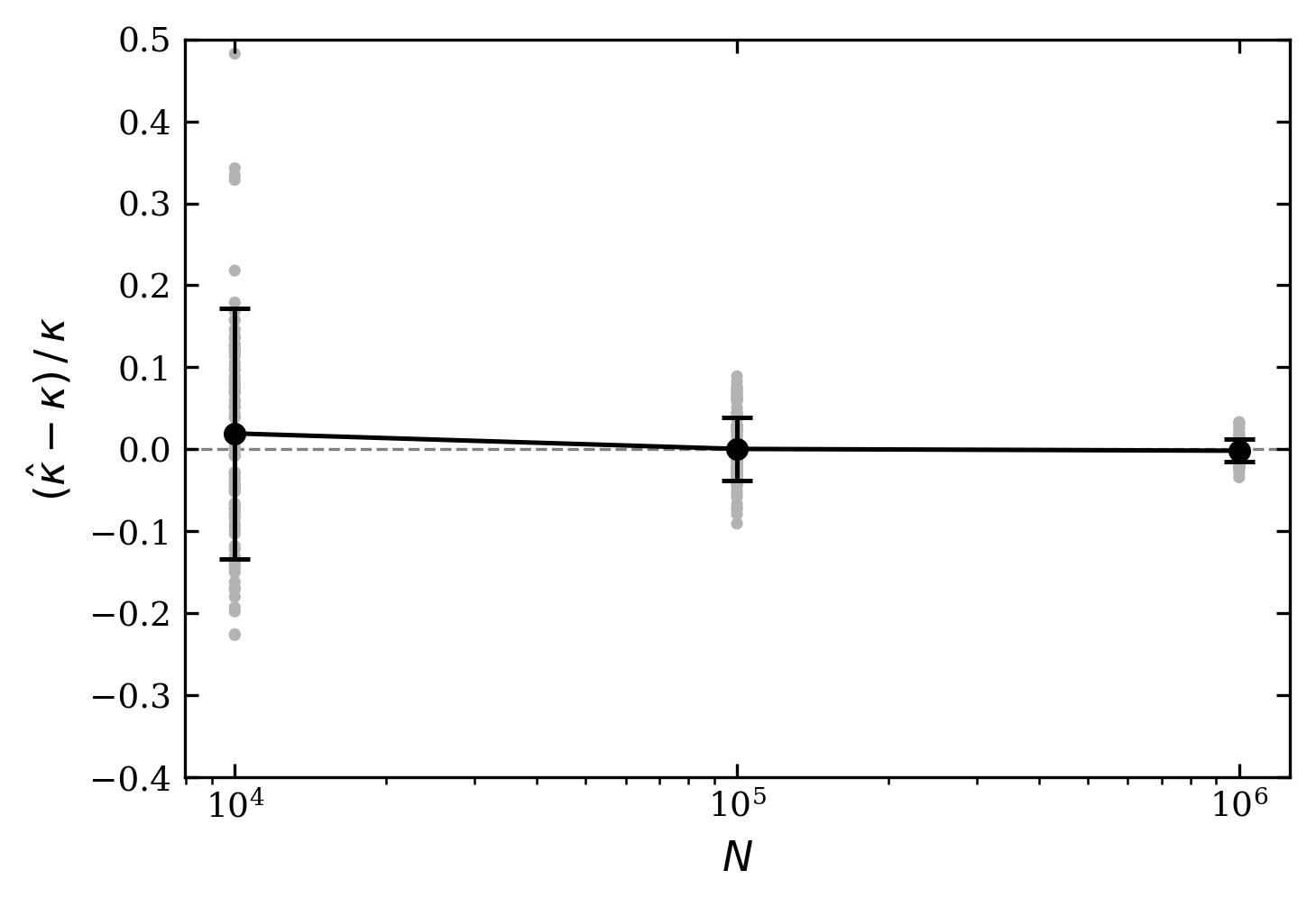}
	\caption{Same as Figure~\ref{fig:error-k2.5}, for \(\kappa = 6.0\).}
	\label{fig:error-k6.0}
\end{figure}

\begin{figure}[H]
	\centering
	\includegraphics[width=0.6\textwidth]{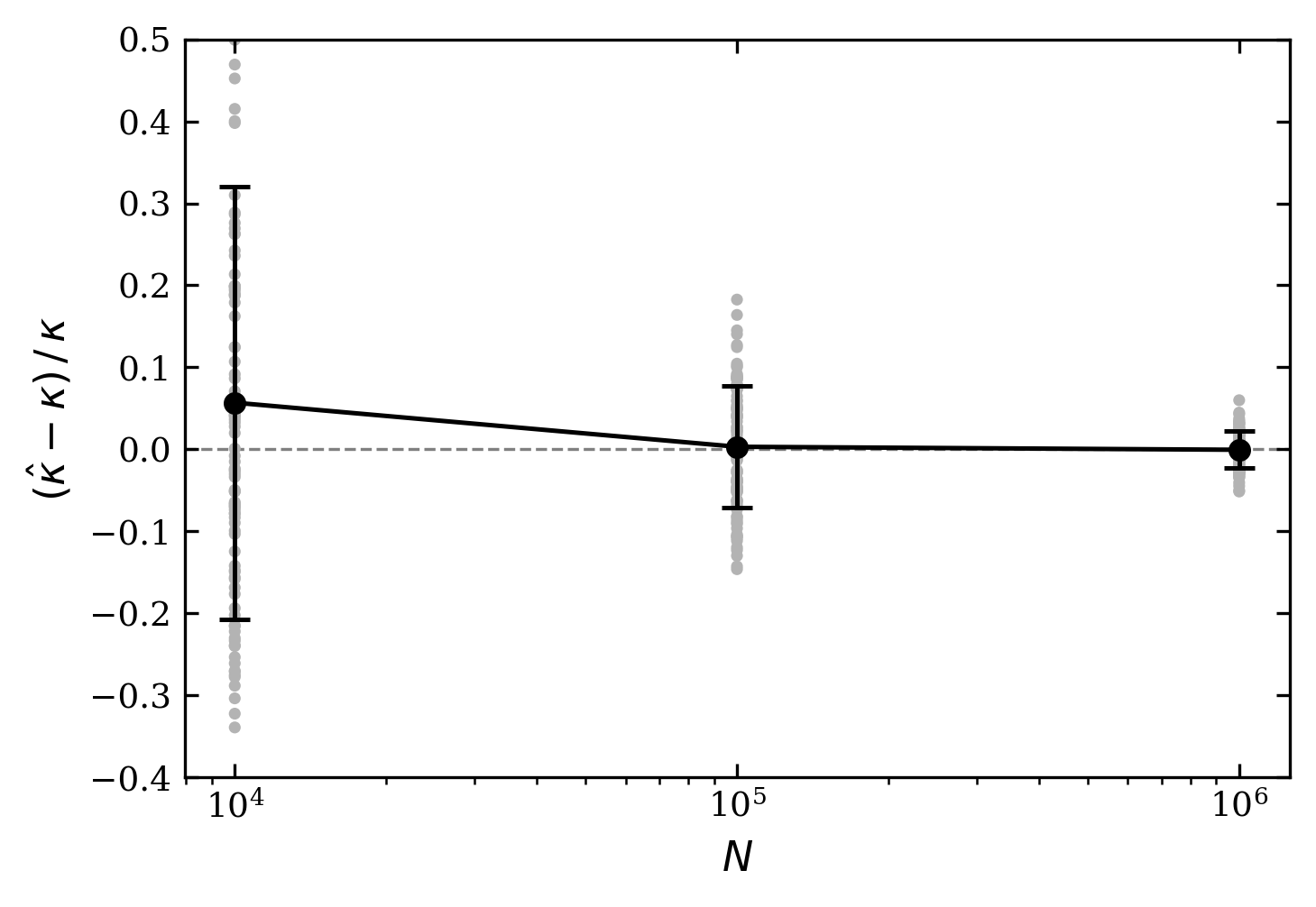}
	\caption{Same as Figure~\ref{fig:error-k2.5}, for \(\kappa = 12.0\).}
	\label{fig:error-k12.0}
\end{figure}

Figures~\ref{fig:error-k2.5},~\ref{fig:error-k6.0} and~\ref{fig:error-k12.0} show the
relative error \((\hat{\kappa} - \kappa)/\kappa\) as a function of the sample size 
\(N\) for \(\kappa = 2.5\), \(6.0\) and \(12.0\), respectively, and 
Tables~\ref{tab:em_results_N1e4}--\ref{tab:em_results_N1e6} report the corresponding 
summary statistics computed over the synthetic data generated.

\vspace{1em}

\begin{table}[H]
	\centering
	\caption{Summary statistics of the EM estimator for sample size \(N = 10^4\).}
	\label{tab:em_results_N1e4}
	\begin{tabular}{c c c c c}
		\hline
		\(\kappa\) & mean(\(\hat{\kappa}\)) & std(\(\hat{\kappa}\)) & mean iter & mean \(t\) (s) \\
		\hline
		2.5  & 2.50  & 0.19 & 378.9  & 0.04 \\
		6.0  & 6.12  & 0.92 & 990.3  & 0.11 \\
		12.0 & 12.68 & 3.17 & 2815.1 & 0.29 \\
		\hline
	\end{tabular}
\end{table}
\vspace{1em}

\begin{table}[H]
	\centering
	\caption{Summary statistics of the EM estimator for sample size \(N = 10^5\).}
	\label{tab:em_results_N1e5}
	\begin{tabular}{c c c c c}
		\hline
		\(\kappa\) & mean(\(\hat{\kappa}\)) & std(\(\hat{\kappa}\)) & mean iter & mean \(t\) (s) \\
		\hline
		2.5  & 2.50  & 0.06 & 354.8  & 0.27 \\
		6.0  & 6.00  & 0.23 & 888.1  & 0.73 \\
		12.0 & 12.04 & 0.89 & 2285.8 & 1.73 \\
		\hline
	\end{tabular}
\end{table}
\vspace{1em}

\begin{table}[H]
	\centering
	\caption{Summary statistics of the EM estimator for sample size \(N = 10^6\).}
	\label{tab:em_results_N1e6}
	\begin{tabular}{c c c c c}
		\hline
		\(\kappa\) & mean(\(\hat{\kappa}\)) & std(\(\hat{\kappa}\)) & mean iter & mean \(t\) (s) \\
		\hline
		2.5  & 2.50  & 0.016 & 335.5  & 3.17  \\
		6.0  & 5.99  & 0.082 & 794.3  & 8.51  \\
		12.0 & 11.99 & 0.270 & 1971.5 & 18.98 \\
		\hline
	\end{tabular}
\end{table}
\vspace{1em}

Three features emerge from the figures and tables. First, \(\text{std}(\hat{\kappa})\) scales as \(N^{-1/2}\) across the explored range; for \(\kappa = 6\), for instance, it decreases from \(0.92\) at \(N = 10^4\) to \(0.08\) at \(N = 10^6\), close to the expected reduction factor of \(\sqrt{N_2/N_1} =\sqrt{10^6/10^4} = 10\), in agreement with the consistency expected of a maximum-likelihood estimator. Second, a small finite-sample bias is visible at \(N = 10^4\) and grows in magnitude with \(\kappa\): it is essentially zero for \(\kappa = 2.5\) and reaches \(\langle\hat{\kappa}\rangle = 12.68\) against \(\kappa = 12.0\) (a relative bias of \(\sim 5.7\%\)) for \(\kappa = 12.0\). For \(\kappa = 6.0\) and \(\kappa = 12.0\) the bias is positive at \(N = 10^4\), changes sign as \(N\) grows, and essentially vanishes by \(N = 10^5\), consistent with the standard \(\mathcal{O}(1/N)\) decay of the bias of maximum-likelihood estimators \cite{lehmann1998theory}. Third, the dispersion of \(\hat{\kappa}\) at fixed \(N\) increases markedly with \(\kappa\): at \(N = 10^4\), \(\text{std}(\hat{\kappa})\) grows from \(0.19\) at \(\kappa = 2.5\) to \(3.17\) at \(\kappa = 12.0\). This reflects the progressive degeneracy of the kappa likelihood as \(\kappa \to \infty\), where the distribution approaches the Maxwell-Boltzmann limit and \(\kappa\) becomes increasingly difficult to identify from a finite sample. The mean iteration count displays a complementary trend: it grows with \(\kappa\), from \(\sim 370\) at \(\kappa = 2.5\) to \(\sim 2800\) at \(\kappa = 12\) for \(N = 10^4\) and decreases mildly with \(N\) at fixed \(\kappa\), as larger samples yield a more concentrated likelihood and accelerate convergence.


\subsection{Discussion}

The superstatistical EM algorithm proposed here provides a statistically rigorous procedure for estimating the spectral index $\kappa$ from velocity data within a hierarchical probabilistic model. By treating the inverse temperature $\beta$ as a gamma-distributed latent variable, the complete-data likelihood acquires exponential-family structure, and both the E-step and the M-step admit analytically closed forms. The kappa distribution itself emerges as the marginal density $P(\bm{v}|\alpha, \theta)$, so the method is internally consistent with the superstatistical interpretation. A central methodological feature is that the framework remains agnostic with respect to the physical interpretation of $\beta$: the algorithm requires only the latent-variable structure and the gamma prior $P(\beta|\alpha, \theta)$, without committing to any specific underlying picture. This addresses a practical gap, since routine estimation of $\kappa$ in plasma physics has often relied on heuristic fitting of histograms without an explicit probabilistic model for the parameter being estimated.

The Monte Carlo study reported in Section~\ref{sec:error-vs-N} confirms the properties expected of a properly derived maximum-likelihood estimator. The standard deviation of $\hat{\kappa}$ decreases as $N^{-1/2}$ at every value of 
$\kappa$, in agreement with consistency, and the small finite-sample bias visible at $N = 10^4$ decays at the standard $1/N$ rate of MLE, vanishing by $N = 10^5$ within the resolution of our experiments~\cite{bishop2006pattern}. 
The log-likelihood increases monotonically along every EM trajectory, providing an internal consistency check on the implementation. Together with the closed-form M-step, these properties make the algorithm both numerically robust 
and theoretically transparent.

The most informative pattern in Tables~\ref{tab:em_results_N1e4} to \ref{tab:em_results_N1e6} is that both 
$\text{std}(\hat{\kappa})$ and the mean iteration count grow with $\kappa$. This behavior is not a defect of the algorithm but a property of the model itself. As $\kappa \to \infty$, the kappa distribution approaches a Maxwellian and the likelihood becomes increasingly flat in $\kappa$, reducing the information per observation. The same conclusion is transparent in the $(u, \beta_S)$ parameterization introduced in Section~\ref{sec:superstatistics}: in 
the Maxwellian limit $\alpha \to \infty$ one has $u \to 0$, so the gamma distribution over $\beta$ concentrates on its mean $\beta_S$ and the data become indistinguishable from those generated under a single fixed inverse temperature. The signal that distinguishes a finite $\kappa$ from the Maxwellian limit simply vanishes, and no estimator, EM or otherwise, can recover it.

The moment-based initialization of Section~\ref{sec:Parameter_initialization} provides a fast, data-driven first estimate of $\alpha$ without iterative computation. 
Its admissibility condition $R > 3$ in one dimension is equivalent to the existence of the fourth velocity moment, that is, $\alpha > 2$ or $\kappa > 3/2$. When this condition fails, either because the sample is nearly Maxwellian and $R$ approaches its lower bound, or because $\kappa$ lies near the boundary of existence, the algorithm reverts to the default initialization $\kappa_0 = 6$, chosen as the midpoint of the range $\kappa \in (2, 10)$ typically reported in space plasmas~\cite{livadiotis2013}. 
This fallback affects only the starting point of the iteration: convergence to the same stationary point of the observed-data likelihood is preserved, although the number of iterations required may increase. A robust alternative based on quantiles or median statistics, well-defined even when $\langle v^4\rangle$ diverges, is a natural extension for the heavy-tail regime $\kappa \to 3/2$ and is left as future work. Further directions include validation on real 
plasma measurements rather than synthetic data, and the extension of the framework to settings where the latent variables $\{\beta_i\}$ are correlated instead of independent, as briefly anticipated in Section~\ref{sec:EM_algorithm_SS}.

\section*{Acknowledgments}

This work was supported by ANID through Beca de Doctorado Nacional 2023 No.~21231822 (L.~H-F.) and FONDECYT grant 1220651 (S.~D.).
\section*{References}
\bibliography{Bibliografia}
\bibliographystyle{unsrt}

\end{document}